\documentclass[preprint,12pt]{elsarticle}

\usepackage{graphicx}
\usepackage{amssymb}

\usepackage{lineno}
\usepackage{amsmath,amsthm,amssymb,amsfonts}
\usepackage{amsfonts}
\usepackage{graphics}
\usepackage{amsmath}
\usepackage{color, epstopdf}
\usepackage{mathrsfs}
\usepackage{graphicx, color, epstopdf}
\usepackage{threeparttable}
\usepackage{float}
\usepackage{array}
\usepackage{subcaption}

\journal{Journal Name}

\begin{document}

\begin{frontmatter}


\title{Learning and Meta-Learning of Stochastic Advection-Diffusion-Reaction Systems from Sparse Measurements}

\author[Hust,Brown]{Xiaoli Chen}
\author[IIT]{Jinqiao Duan}
\author[Brown,PNNL]{George Em Karniadakis\corref{cor}}
\cortext[cor]{Corresponding author}
\ead[cor]{george\_karniadakis@brown.edu}
\address[Hust]{Center for Mathematical Sciences \& School of Mathematics and Statistics, Huazhong
University of Science and Technology, Wuhan 430074, China}
\address[Brown]{Division of Applied Mathematics, Brown University, Providence, RI 02912, USA}
\address[IIT]{Department of Applied Mathematics, Illinois Institute of Technology, Chicago, IL 60616, USA}
\address[PNNL]{Pacific Northwest National Laboratory, Richland, WA 99354, USA}

\begin{abstract}
Physics-informed neural networks (PINNs) were recently proposed in~\cite{raissi2019physics} as an alternative
way to solve partial differential equations (PDEs). A neural network (NN) represents the solution while a PDE-induced NN
is coupled to the solution NN, and all differential operators are treated using automatic differentiation.
Here, we first employ the standard PINN and a stochastic version, sPINN, to solve forward and inverse problems governed by a nonlinear advection-diffusion-reaction (ADR) equation, assuming we have some sparse measurements of the concentration
field at random or pre-selected locations. Subsequently,  we attempt to optimize the hyper-parameters of sPINN by using
the Bayesian optimization method (meta-learning), and compare the results with
the empirically selected hyper-parameters of sPINN.
In particular, for the first part in solving the inverse deterministic ADR, we assume that
we only have a few high-fidelity measurements whereas the rest of the data is of lower fidelity. Hence, the PINN is trained using a composite {\em multi-fidelity} network, first introduced in~\cite{Meng2019}, that learns the correlations between the multi-fidelity data and predicts the unknown values of diffusivity, transport velocity, and two reaction constants as well as the concentration field.
For the stochastic ADR, we employ a Karhunen-Lo\`{e}ve (KL) expansion to represent the stochastic diffusivity, and
arbitrary polynomial chaos (aPC) to represent the stochastic solution. Correspondingly, we design multiple NNs
to represent the {\em mean} of the solution and learn each aPC mode separately whereas we employ a separate NN to
represent the {\em mean} of diffusivity and another NN to learn all modes of the KL expansion. For the inverse
problem, in addition to stochastic diffusivity and concentration fields, we also aim to obtain the (unknown) deterministic values of transport velocity and reaction constants. The available
data correspond to seven {\em spatial} points for the diffusivity and 20 {\em space-time} points for the solution, both sampled 2,000 times.
We obtain good accuracy for the deterministic parameters of the order of $1\% - 2\%$, and excellent accuracy for the mean and variance of
the stochastic fields, better than three digits of accuracy.
In the second part, we consider the previous stochastic inverse problem and we use Bayesian optimization to find
five hyper-parameters of sPINN, namely the width, depth and learning rate of two NNs for learning the modes. We obtain much deeper and wider optimal NNs compared to the
manual tuning, leading to even better accuracy, i.e., errors less than $1\%$ for the deterministic values, and about an order of
magnitude less for the stochastic fields.
\end{abstract}


\end{frontmatter}


\section{\label{Introduction}Introduction}
In classical inverse problems we assume that we have a lot of measurements for the state variables, and we aim to obtain some
unknown parameters or space/time-dependent material properties by formulating appropriate objective functions and employing
the necessary regularization techniques. However, in many practical problems, e.g., in subsurface transport~\cite{Tartakovsky,Tartakovsky1}, we
have to deal with a {\em mixed} problem, as we typically have some measurements on the material properties and some measurements on
the state variables. Here, we consider such ``mixed" problems for a nonlinear advection-diffusion-reaction (ADR)
describing a concentration field,
and we formulate new algorithms inspired by recent developments in machine learning. In particular, we will assume that
we have a stochastic diffusivity field, which is partially known only at a few points, and hence we aim to determine the entire stochastic field
from only sparse measurements of the concentration field. Moreover, we will assume that the constant
transport velocity in the advection term is unknown and that the reaction term is parametrized by two unknown parameters. Hence, the problem set up we consider is as follows: determine the entire stochastic diffusivity and stochastic concentration fields as well three (deterministic) parameters from a few multi-fidelity measurements of the concentration field at random points in space-time. For simplicity we will refer to this ``mixed" problem as ``inverse" problem in the following.

The aforementioned problem set up could be tackled by using Bayesian optimization methods as we have done in previous work for other problems,
e.g., see \cite{raissi2018numerical,pang2019neural}, but to overcome open issues related to strong nonlinearity and scalability, here we will employ neural networks (NNs) following the works of \cite{han2018solving,chen2018neural,chaudhari2018deep,sirignano2018dgm,qin2019data},
and in particular the physics-informed neural network (PINN) approach introduced in \cite{raissi2019physics}. In addition, we have to model stochastic fields and in order to avoid optimizing expensive Bayesian NNs, we will instead model stochasticity using polynomial
chaos expansions following the work of \cite{zhang2019quantifying}. Another important consideration is how to fuse data of variable fidelity,
as some data may be collected by a few high-resolution sensors whereas the majority of the data may be collected by lower fidelity sensors. This, in turn, implies that we have to train the NN or the PINN with multi-fidelity data, and to this end we will employ a new composite network recently proposed by~\cite{Meng2019}. Finally, because of the complexity of the proposed NNs,
we also introduce an automated method to optimize the hyper-parameters of PINN using a simple version of meta-learning, i.e., Bayesian
optimization, e.g., see \cite{bergstra2011algorithms,snoek2012practical,snoek2015scalable}.

In order to make progress towards the final goal and to evaluate each of the algorithmic steps separately, we will use a hierarchical approach
by introducing complexity incrementally. We will start with multi-fidelity deterministic problems using PINNs and subsequently we will introduce randomness in the data and present the stochastic formulation. This will require us to design multiple NNs that learn in modal space.
Subsequently, we will formulate an additional optimization problem for five of the most important hyper-parameters of the multi-NN design, and compare its performance with the performance obtained previously by manual tuning.

The organization of this paper is as follows. In Section \ref{Methodology}, we introduce the PINN to solve the deterministic partial differential equation and the sPINN to solve the stochastic partial differential equation. In Section \ref{Deter}, we present the results of PINN for solving the inverse problem of the deterministic ADR equation. In Section \ref{Stoc}, we provide the results of the sPINN method for both the forward and inverse problem. Finally,  we employ meta-learning for the last stochastic inverse problem, and we conclude with a short summary.


\section{\label{Methodology}Methodology}
\subsection{PINNs: Physics-informed neural networks for deterministic PDEs}
First, we briefly review the type of deep neural networks (DNNs) to solve deterministic partial differential equations (PDEs) and the corresponding inverse problem \cite{raissi2019physics, pang2019fpinns}. The PDE can have the general form:
\begin{align}
u_t+\mathscr{N}[u(x,t);\eta]=0, ~x \in \mathcal{D}, ~t\in[0,T],\label{deter-model}
\end{align}
with the initial and boundary conditions:
\begin{align}
&u(x,0)=u_0(x),~~~~~~~~~~~~~~x \in \mathcal{D},\nonumber\\
&\mathbb{B}_X[u(x,t)]=\tilde{u}(x,t),~~~~~~~x \in \partial \mathcal{D},~t\in (0,T],\
\end{align}
where $u(x,t)$ denotes the solution, $u_0(x)$ is the initial condition, $\tilde{u}(x,t)$ is the boundary condition, $\mathscr{N}[\cdot]$ is a nonlinear differential operator, $\eta$ is the parameter in the PDE, $\mathcal{D}$ is a subset of $\mathbb{R}$, and $\partial\mathcal{D}$ is the boundary of $\mathcal{D}$.

\begin{figure}[ht]
\centerline{\includegraphics[width=10cm,height=4cm]{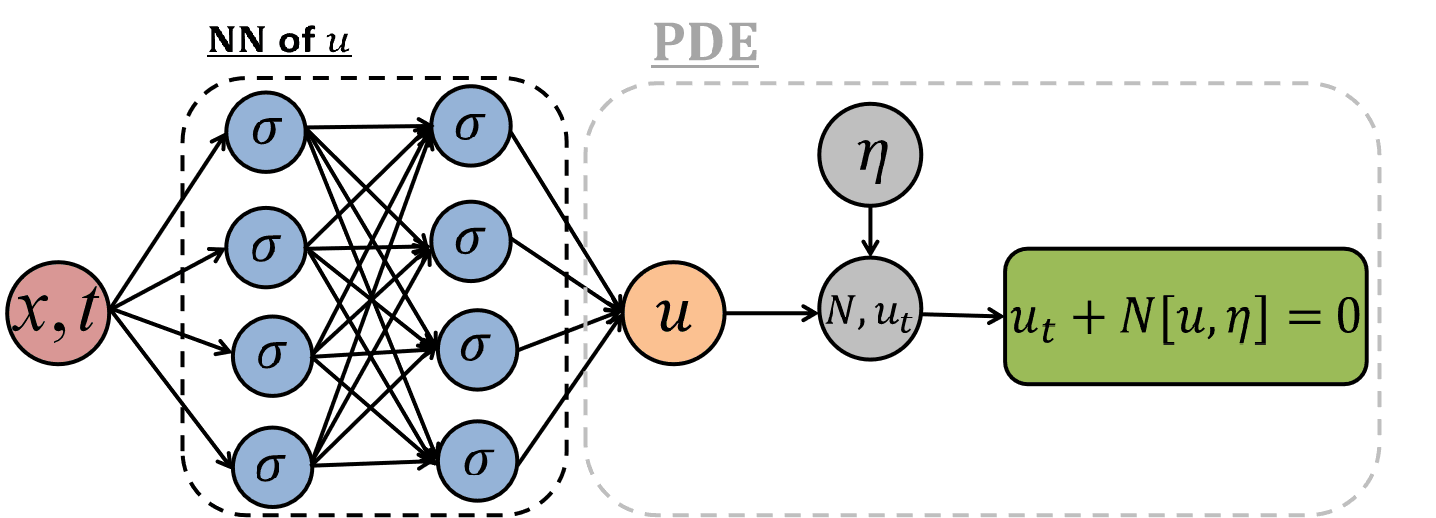}}
\caption{\textbf{} Schematic of the PINN for solving deterministic partial differential equations. }
\label{NN1}
\end{figure}

The solution, denoted by $u_{NN}(x,t; w,b)$, is constructed as a neural network approximation of $u(x,t)$; DNN
has the weights ($w$) and biases ($b$). We can couple it to another DNN induced by the PDE residual $f_{NN}$ computed based on the NN solution $u_{NN}(x,t; w,b)$ and corresponding to Equation \eqref{deter-model}; also, the residual $f(x,t)$ is given by  Equation \eqref{deter-model}, i.e.,
\begin{align}
f=u_t+\mathscr{N}[u(x,t);\eta].
\label{deter-model-f}
\end{align}
The inputs of the DNN are the spatial coordinates and time $(x,t)$ while the output is $u_{NN}$, which has the same dimension as the input.
For the output of $u_{NN}$, we use automatic differentiation techniques to compute all derivatives of the nonlinear differential operator (physics part).
There are two restrictions on $u_{NN}$. First, the solution of $u_{NN}$ should be close to the observations $u$ at the training points. Second, every $u_{NN}$ should comply with the physics imposed by Equation \eqref{deter-model}. The second part is achieved by defining a residual network:
\begin{align}
f_{NN}(x,t;w,b,\eta)=(u_{NN})_t+\mathscr{N}[u_{NN}(x,t;w,b);\eta],
\end{align}
which is computed from $u_{NN}$ straightforwardly with automatic differentiation.
This residual network network $f_{NN}$, shares the same parameters $(w,b)$ with the network for $u_{NN}$ and should output a value close to $0$ for any input $(x,t)\in\mathcal{D} \times [0,T]$.
During training, the shared parameters $(w,b)$ are adjusted by back-propagating the error obtained by minimizing a loss function that is the weighted sum of the above two constraints.
A sketch of the PINN, consisted of the physics-uninformed and physics-informed DNNs is shown in Fig. \ref{NN1}.

The PINN loss function is defined as:
\begin{align}
MSE=MSE_u+MSE_f,
\end{align}
where
\begin{align}
MSE_u&=\frac{1}{N_u} \sum_{i=1}^{N_u} (u_{NN}(x_i,t_i;w,b) -u(x_i,t_i))^2, \nonumber\\
MSE_f&=\frac{1}{N_f} \sum_{j=1}^{N_f} (f_{NN}(x_j,t_j; w,b,\eta) )^2.\
\end{align}
Here, $(x_i,t_i,u_{NN}(x_i,t_i;w,b))_{i=1}^{N_u}$ denote the initial and boundary conditions of $u$ for
the forward problem as well as the training data of $u$ for the inverse problem. The data $u(x_i,t_i)$ are the observation data of $u$ while
 $\{(x_j,t_j)\}_{j=1}^{N_f}$ denote the residual points for penalizing $f(x,t)$.

\subsection{sPINNs: Physics-informed neural networks for stochastic PDEs}
Next, we briefly review a stochastic version, based on the arbitrary polynomial chaos \cite{XiaoliangWan_GK,paulson2017arbitrary}
to represent stochasticity and combine it with a PINN, following the method first introduced in~\cite{zhang2019quantifying}.
We consider the following stochastic PDE (SPDE):
\begin{align}
u_t+\mathscr{N}[u(x,t;\omega);k(x;\omega)]=0, ~x \in \mathcal{D}, ~t\in(0,T],~\omega\in \Omega ,\label{sto-model}
\end{align}
with the initial and boundary conditions:
\begin{align}
&u(x,0;\omega)=u_0(x),~~~~~~~~x \in \mathcal{D},\nonumber\\
&\mathbb{B}_X[u(x,t;\omega)]=0,~~~~~~~~x \in \partial\mathcal{D},~t\in(0,T].\
\end{align}

Here $\Omega$ is the random space. In the following, we describe how to use sPINN to solve stochastic inverse problems since for the forward problem the method is straightforward.
We assume that we have $N_k$ sensors for $k(x;\omega)$ placed at $\{x_k^{(i)}\}_{i=1}^{N_k}$ and $N_u$ sensors for $u(x,t;\omega)$ placed at $\{(x_u^{(i)},t_u^{(i)})\}_{i=1}^{N_u}$. We also choose at random $N_f$ locations $\{(x_f^{(i)},t_f^{(i)})\}_{i=1}^{N_f}$ that are used to compute the residual of Equation \eqref{sto-model}. We assume that the observation data of $k$ are $\{k(x_i;\omega_s)\}$ (denoted by $\{k_s^{i}\}$ ), where $i=1,2,...,N_k$, and $s=1,2,...,N$.
The observation data of $u$ is $\{u(x_j,t_j;\omega_s)\}$ (denoted by $\{u_s^{j}\}$ ), where $j=1,2,...,N_u$, and $s=1,2,...,N$.
Here $N$ denotes the number of samples available for a specific location, and for simplicity we take
that to be the same both for $k(x;\omega)$ and for $u(x,t;\omega)$ for all locations.

One of the key questions for the inverse stochastic problem is what type of randomness we encounter in the data and how we represent the stochastic fields. We consider a general setting, i.e., instead of the classical inverse problem where we are given data on $u(x,t;\omega)$ but not on $k(x;\omega)$, here we assume that we have some data on $u$ and some data on $k$. Hence, in order to choose the type of the distribution required to represent our random variables so that we employ arbitrary polynomial chaos (aPC), we use the data samples of either $u$ or $k$. Here we assume that we have $N_k=7$ sensors for $k$ so we can determine the random variables $\xi$ from the k-data, as we explain below.

We choose $M$ sensors of $k$ to compute the random variables $\xi$, where $M \leq N_k$. Denote the observations of $k$ as $k_1=(k_1(i,j))$, where the element of $k_1(i,j)$ is the value of $k(x_i;\omega_j)$, and the size of $k_1$ is $M \times N$, where $N$ is the number of samples. $K$ be the $M \times M$ covariance matrix for the observation data of $k_1$, i.e.
  \begin{align}
K_{i,j}=Cov(k_1^{(i)},k_1^{(j)}).
\end{align}
Let $\lambda_i$ and $\upsilon_i$ be the i-th eigenvalue and its corresponding normalized eigenvector of $K$. Using principal component analysis (PCA) we obtain
\begin{align}
K=V^T \Lambda V,
\end{align}
where $V=[\upsilon_1, \upsilon_2,...,\upsilon_{M}]$ is an orthonormal matrix and $\Lambda=diag(\lambda_1,\lambda_2,...,\\
\lambda_{M})$ is a diagonal matrix.
The random variable $\xi$ satisfies the following equation
\begin{align}
k_1=\bar{k_1}+V \sqrt{\Lambda} \xi,
\end{align}
where $\bar{k_1}$ is the mean of $k_1$.

Hence
\begin{align}
\xi= \sqrt{\Lambda}^{-1} V^T (k_1-\bar{k_1}),
\end{align}
where each row of $\xi$ is an uncorrelated random vector, and the size of $\xi$ is $M \times N$.

In the continuous case, the diffusion term $k(x;\omega)$ can be approximated by:
\begin{align}
k_{NN}(x;\omega_j)= k_0(x)+\sum_{i=1}^{M}k_i(x) \sqrt{\lambda_i}\xi_{i,j},~~j=1,...,N.\label{KL3}
\end{align}
Correspondingly, the solution $u$ at the $j$-th snapshot can be approximated by
\begin{align}
u_{NN}(x,t;\omega_j)\approx\sum_{\alpha=0}^{P}u_{\alpha}(x,t)\psi_{\alpha}(\xi_j),
\end{align}
%
%
where $\{\psi_\alpha\}_{\alpha=1}^P$ are the set of multivariate orthonormal polynomial basis and the highest polynomial order is $r$. The parameter $P$, $r$ and $M$ satisfy the following formula
\begin{align}
P+1=\frac{(r+M)!}{r!M!}.
\end{align}

\begin{figure}[h]
\begin{minipage}[]{0.5 \textwidth}
\centerline{\includegraphics[width=6cm,height=3cm]{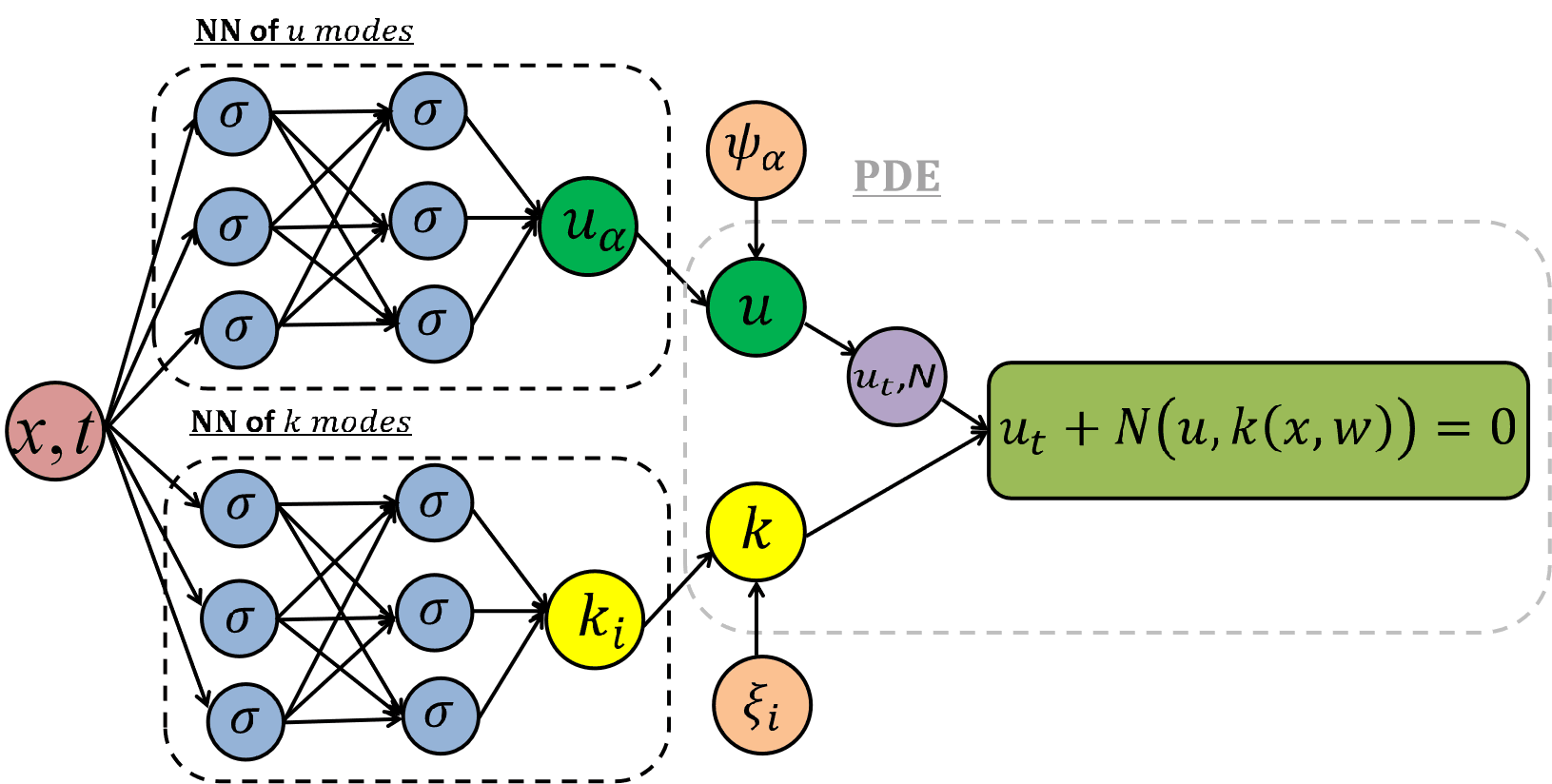}}
\end{minipage}
\begin{minipage}[]{0.5 \textwidth}
\centerline{\includegraphics[width=4.5cm,height=3cm]{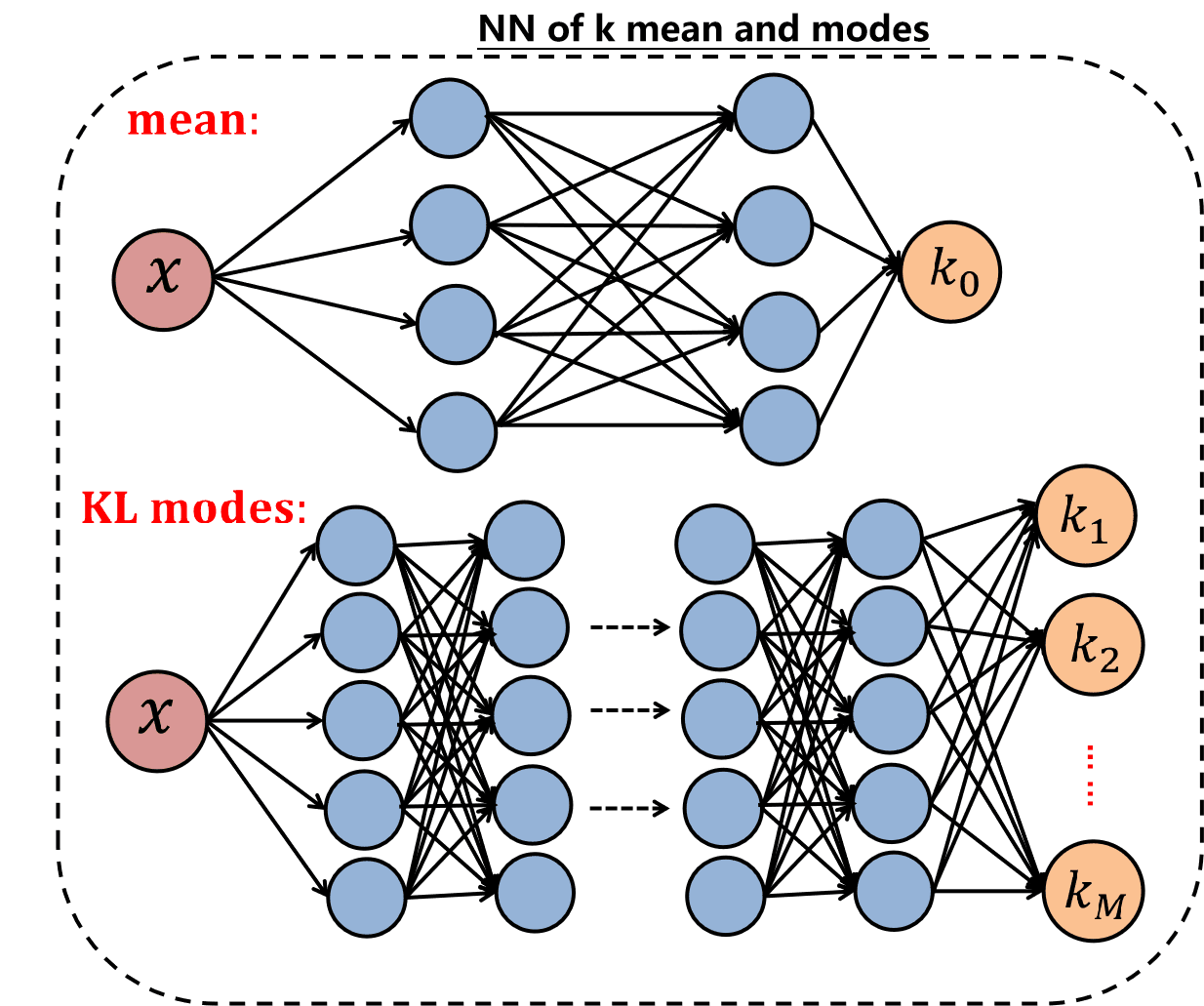}}
\end{minipage}
\begin{minipage}[]{0.5 \textwidth}
\centerline{\includegraphics[width=6cm,height=3cm]{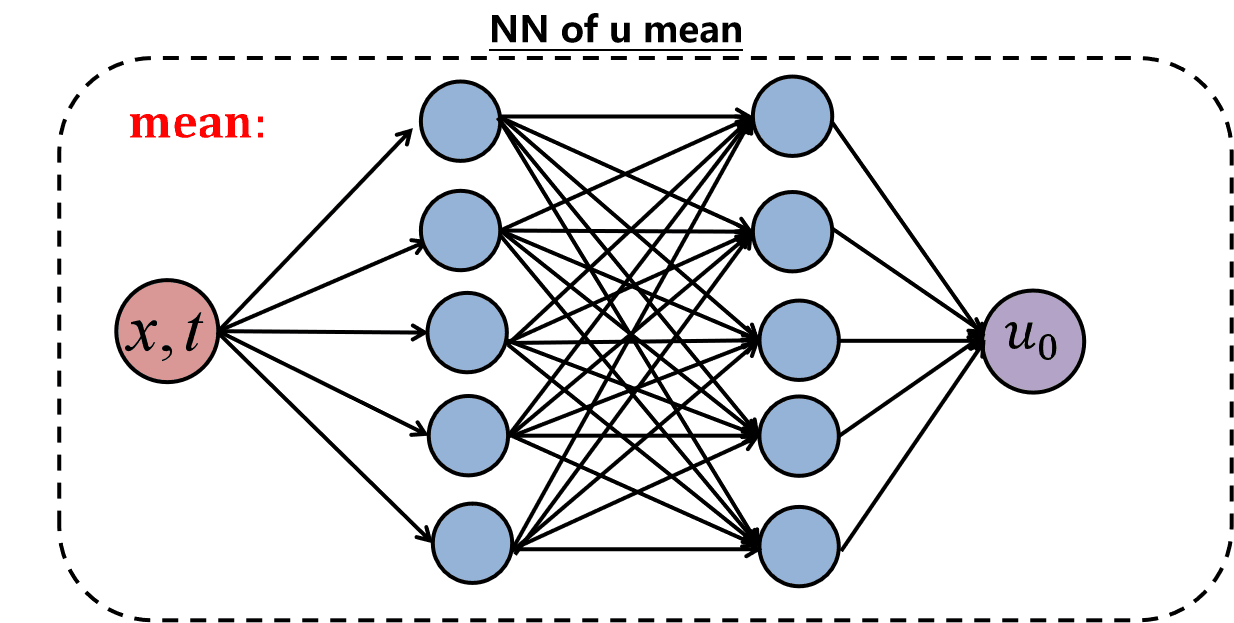}}
\end{minipage}
\begin{minipage}[]{0.5 \textwidth}
\centerline{\includegraphics[width=6cm,height=3cm]{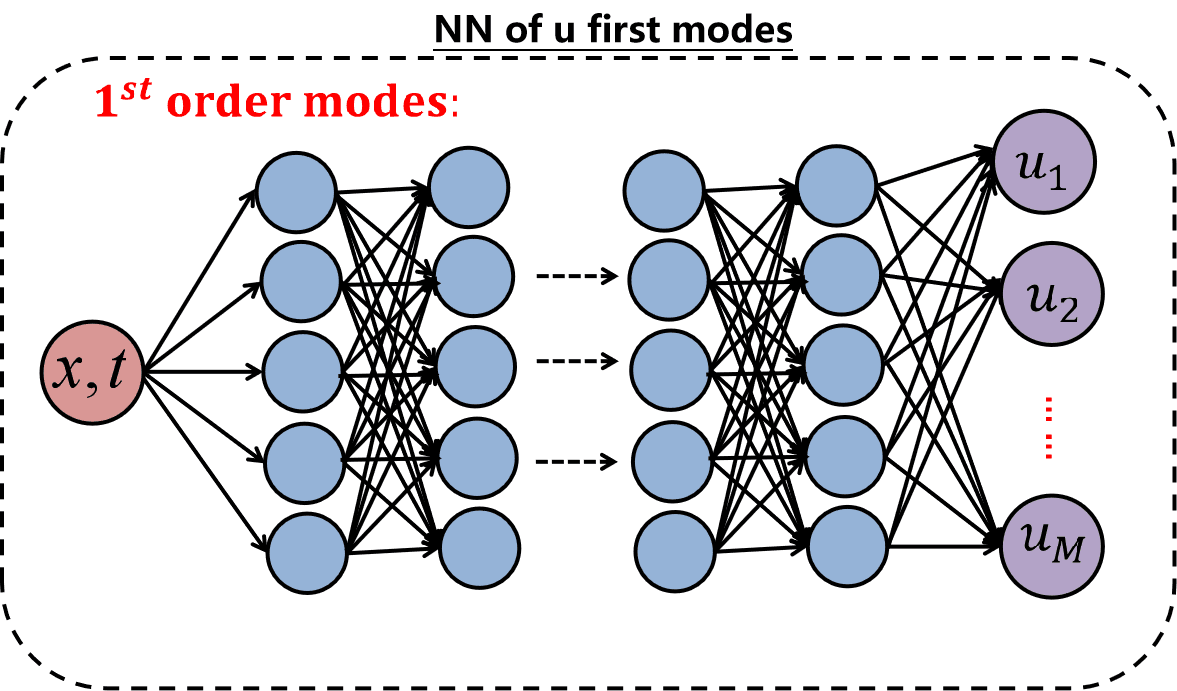}}
\end{minipage}
\begin{minipage}[]{0.5 \textwidth}
\centerline{\includegraphics[width=6cm,height=3cm]{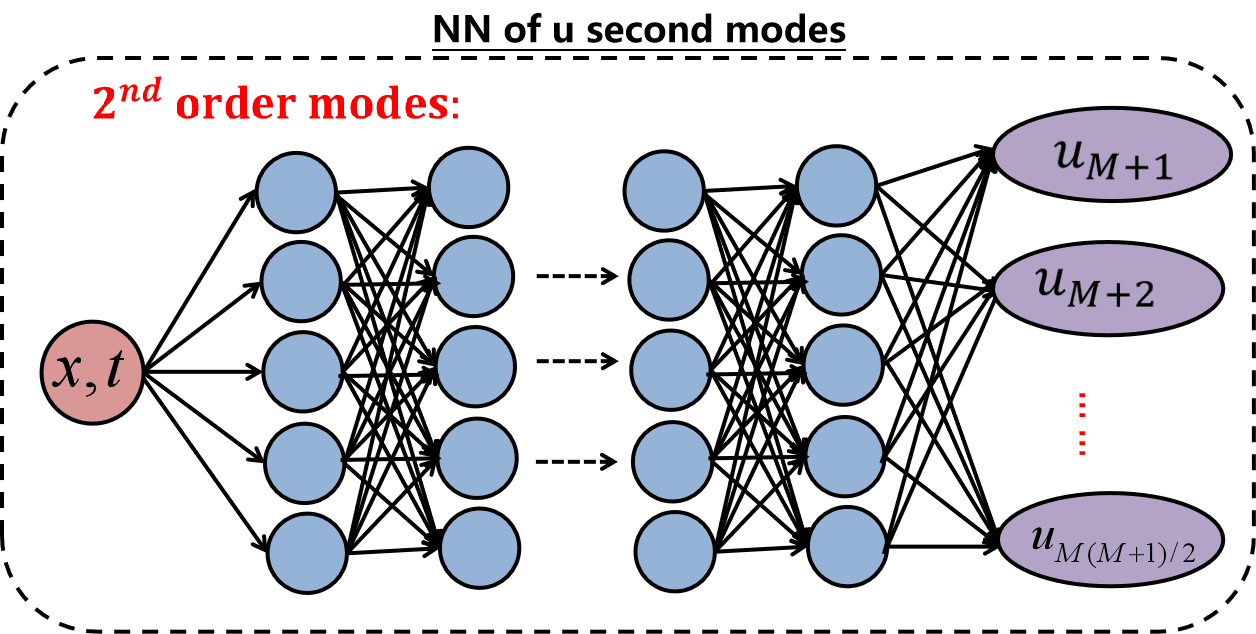}}
\end{minipage}
\begin{minipage}[]{0.5 \textwidth}
\centerline{\includegraphics[width=6cm,height=3cm]{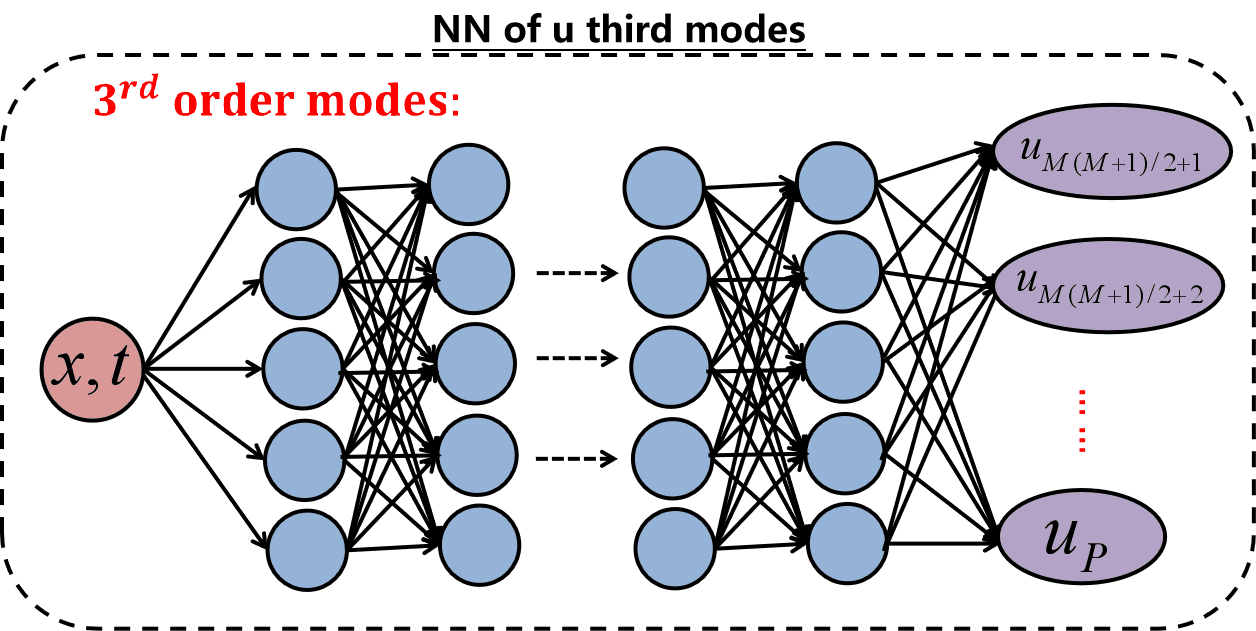}}
\end{minipage}
\caption{\textbf{}Schematic of the sPINN for solving stochastic partial differential equations. Top left:
A composite NN consisting of multiple NNs for computing the mean and modes of the
stochastic diffusivity and the solution. Top right: Two separate NN for the mean and all the modes of diffusivity.
Middle left: NN of to compute the mean of the solution. Middle right and bottom row: Separate NN to compute the modes of
the solution. Adopted from reference \cite{zhang2019quantifying}. }
\label{NNaPC}
\end{figure}

Similar to the PINN method, we construct the residual network via automatic differentiation and by substituting $u(t,x; \omega)$ and $k(x,\omega)$ in Equation \eqref{sto-model} with $u_{NN}(x,t;\omega)$ and $k_{NN}(x;\omega)$.
A sketch of the stochastic PINN (sPINN) is shown in Fig. \ref{NNaPC}.

The loss function is defined as:

\begin{align}
MSE=MSE_{u}+MSE_{k}+MSE_{f},
\end{align}
where
\begin{align}
MSE_{u}&=\frac{1}{N*N_u}\sum_{j=1}^{N}\sum_{i=1}^{N_u}[u_{NN}(x_u^{(i)},t_u^{(i)};\omega_j)-u(x_u^{(i)},t_u^{(i)};\omega_j)]^2,\nonumber\\
MSE_{k}&=\frac{1}{N*N_k}\sum_{j=1}^{N}\sum_{i=1}^{N_k}[k_{NN}(x_k^{(i)};\omega_j)-k(x_k^{(i)};\omega_j)]^2,\nonumber\\
MSE_{f}&=\frac{1}{N*N_f}\sum_{j=1}^{N}\sum_{i=1}^{N_f}[f_{NN}(x_f^{(i)},t_f^{(i)};\omega_j)]^2.\nonumber\
\end{align}

\section{\label{Deter}Results for the deterministic PDE}
We start with a deterministic PDE to demonstrate how can we infer some of the unknown parameters using PINNs. We will assume that we have different types of data
of variable fidelity, and we will also demonstrate how we can make use of data of lower fidelity as well using the composite neural network (NN) first introduced in~\cite{Meng2019}. We consider the following nonlinear ADR equation:
\begin{align}
\left\{
\begin{array}{ll}
u_t=\nu_1 u_{xx}-\nu_2 u_x+g(u),
& (x,t)\in(0,\pi)\times(0,1], \\
u(x,0)=u_0(x),
& x\in (0,\pi),\\
u(0,t)=1,~u_x(\pi,t)=0,
& t\in(0,1].
\end{array}\right.\label{model}
\end{align}
%
We define the residual $f=u_t-\nu_1 u_{xx}+\nu_2 u_x-g(u)$. The $L_2$ error of a function $h$ is defined as $E_{h}=||h_{NN}-h_{true}||_{L2}$, and the relative $L_2$ error is defined as $E_{h}=\frac{||h_{NN}-h_{true}||_{L2}}{||h_{true}||_{L2}}$.

\subsection{Single-fidelity data}
First, we will use single-fidelity data to infer different parameters and at the same time obtain the solution $u$.
We consider the initial condition $u_0(x)=\exp(-10x)$ and the reaction term $g(u)=\lambda_1 u^{\lambda_2}$. We aim to infer the parameters $\nu_1, \nu_2, \lambda_1, \lambda_2$ given
some sparse measurements of $u$ in addition to initial and boundary conditions. The correct values for the ``unknown" parameters are:
$\nu_1=1, \nu_2=1, \lambda_1=-1, \lambda_2=2$.

We employ the following loss function in the PINN:
\begin{align}
MSE=MSE_u+w_{\nabla_u}*MSE_{\nabla u}+MSE_f,\
\end{align}
where
\begin{align}
MSE_u&=\frac{1}{N_u}\sum_{i=1}^{N_u}|u_{NN}(t_u^i,x_u^i)-u^i|^2,\nonumber\\
MSE_{\nabla u}&=\frac{1}{N_u}\sum_{i=1}^{N_u}|\nabla u_{NN}(t_u^i,x_u^i)-\nabla u^i|^2,\nonumber\\
MSE_f&=\frac{1}{N_f}\sum_{i=1}^{N_f}|f_{NN}(t_f^i,x_f^i)|^2.\nonumber\
\end{align}
The points $\{t_u^i,x_u^i,u_{NN}(t_u^i,x_u^i)\}_{i=1}^{N_u}$ denote the training data for $u(t,x)$,
and ~$N_u=64$,~$N_f=1089$, and $u^i$ is the ``reference solution", which is computed by the second-order finite difference method ($\Delta x=\frac{\pi}{1024}$ and $\Delta t=\frac{1}{1600}$). We use 4 hidden layers and 20 neurons per layer for the deep neural network (DNN).
%
The error of the parameters is defined as $E_{\nu_1}=\frac{\nu_{1train}-\nu_1}{\nu_1}$, $E_{\nu_2}=\frac{\nu_{2train}-\nu_2}{\nu_2}$, $E_{\lambda_1}=\frac{\lambda_{1train}-\lambda_1}{\lambda_1}$ and $E_{\lambda_2}=\frac{\lambda_{2train}-\lambda_2}{\lambda_2}$.

\begin{figure}[ht]
\begin{minipage}[]{0.5 \textwidth}
\centerline{\includegraphics[width=6cm,height=3cm]{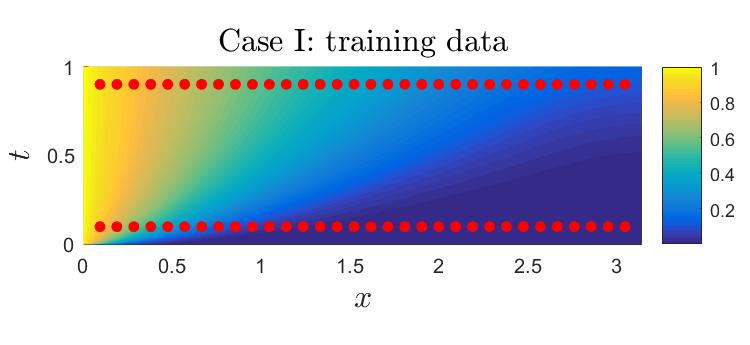}}
\end{minipage}
\begin{minipage}[]{0.5 \textwidth}
\centerline{\includegraphics[width=6cm,height=3cm]{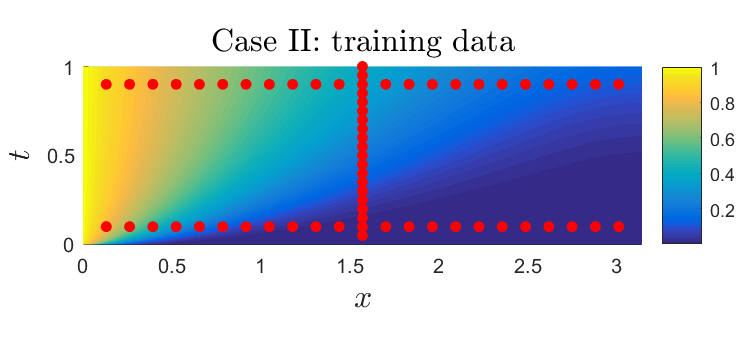}}
\end{minipage}
\begin{minipage}[]{0.5 \textwidth}
\centerline{\includegraphics[width=6cm,height=3cm]{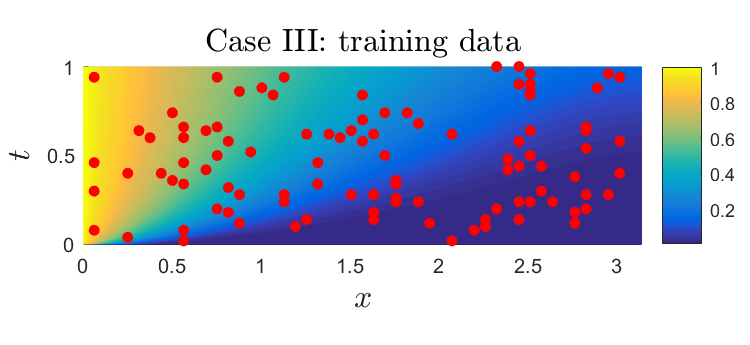}}
\end{minipage}
\begin{minipage}[]{0.5 \textwidth}
\centerline{\includegraphics[width=6cm,height=3cm]{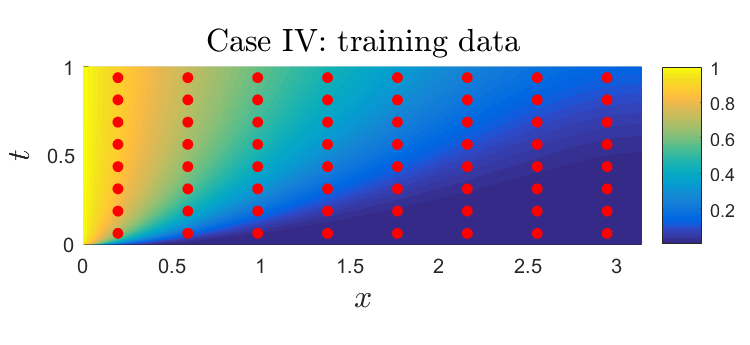}}
\end{minipage}
\caption{\textbf{} Single-fidelity case: The position of training data used in
the loss function. }
\label{f-ptd}
\end{figure}

\begin{figure}[ht]
\begin{minipage}[]{0.5 \textwidth}
\centerline{\includegraphics[width=6cm,height=4cm]{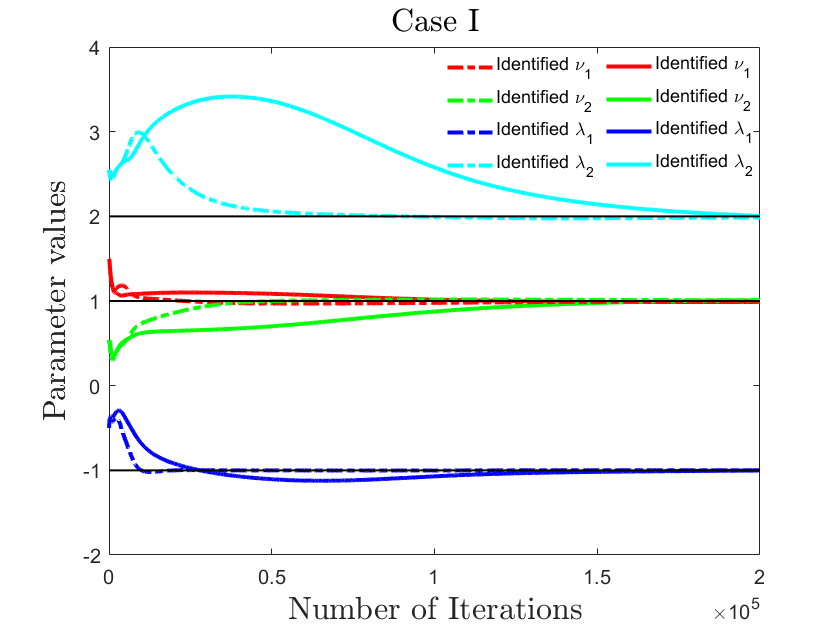}}
\end{minipage}
\begin{minipage}[]{0.5 \textwidth}
\centerline{\includegraphics[width=6cm,height=4cm]{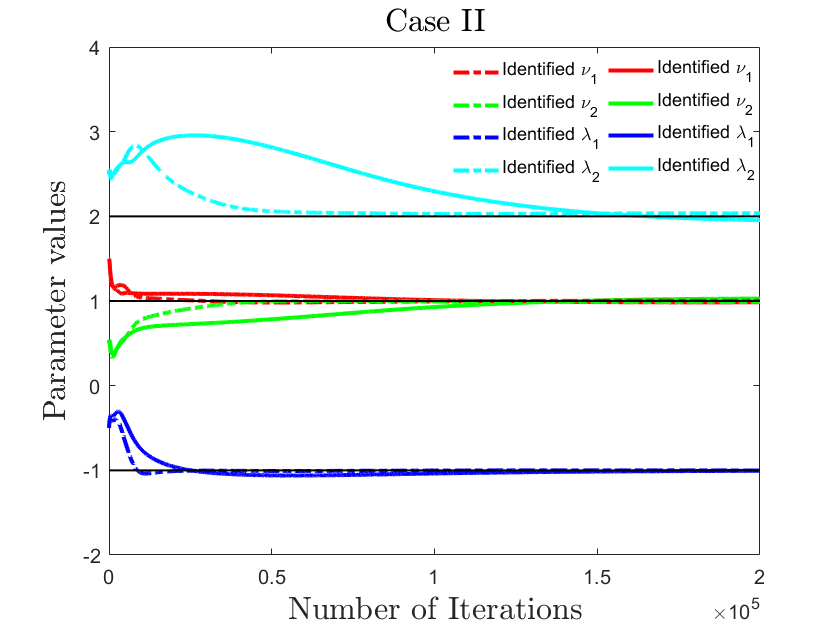}}
\end{minipage}
\begin{minipage}[]{0.5 \textwidth}
\centerline{\includegraphics[width=6cm,height=4cm]{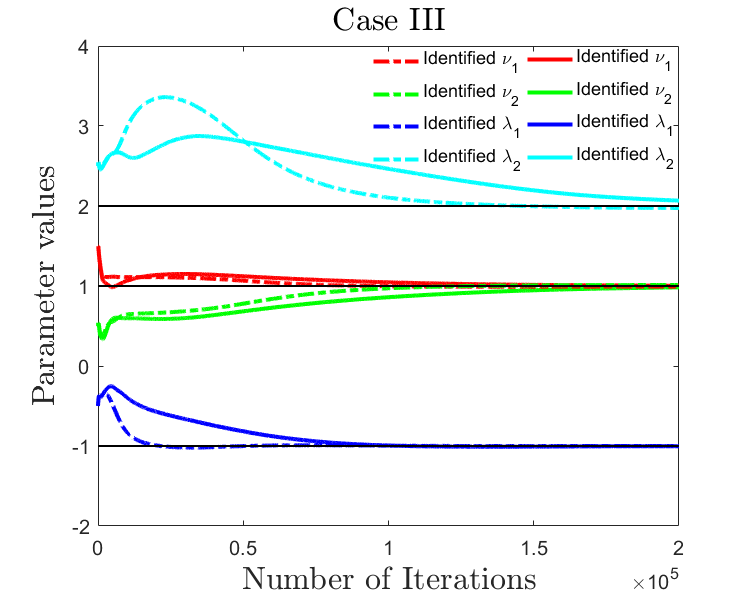}}
\end{minipage}
\begin{minipage}[]{0.5 \textwidth}
\centerline{\includegraphics[width=6cm,height=4cm]{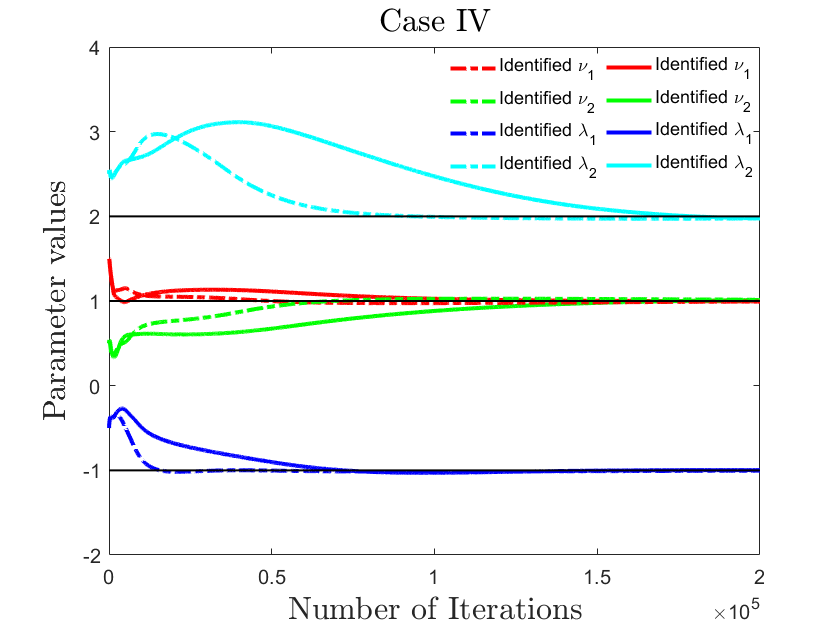}}
\end{minipage}
\caption{\textbf{} Single-fidelity case. Parameter evolution as the iteration of optimizer progresses for four different training data sets. The solid line corresponds to a loss
without penalizing the gradient term while the dash like corresponds
to a loss that includes the gradient term.}
\label{f-para-sig}
\end{figure}

In the following, we will investigate four different ways to choose the training points as shown in Fig. \ref{f-ptd}. For case I, the training data come from two snapshots at $t=0.1$ and $t=0.9$. For case II, the training data come from three snapshots at $t=0.1$, $t=0.9$, and $x=\frac{\pi}{2}$.
For case III, we choose the training data randomly. For case IV, we assume that we have the training data on a regular lattice in the $x-t$ domain.
In all cases we have $64$ training points, and for the weights in the loss function
we investigate both the case with $w_{\nabla u}=0$ and also the case with
$w_{\nabla u}=1$. In the latter case, we assume that we also have available the gradients of the field $u$.
We present the parameter evolution predictions as the iteration of
the optimizer progresses in Fig. \ref{f-para-sig}. The convergence is faster if we include the gradient penalty term ($w_{\nabla_u}=1$ ).
We summarize the results in terms  of the error of the solution $u(x,t)$ and of the parameters in Table \ref{tab:sig}.

\begin{table*}[ht]
\scriptsize
\begin{center}
\caption{Single-fidelity case: Errors of the solution and of the parameters.}
\begin{tabular}{ c cc cc cc cc cc c cc cc cc cc ccc c}
\hline
                & $E_u$         &$E_{\nu_1}     $& $E_{\nu_2} $&$E_{\lambda_1}$ & $E_{\lambda_2}$ \\[1ex]
\hline
Case I ($w_{\nabla u}=0$)  & $1.9514e-03$&$1.1098$ \%& $ 0.9010$ \% &$0.2240$ \% &$ 0.2370$ \%  \\[1ex]
Case I ($w_{\nabla u}=1$)  &$1.6405e-03 $&$0.8004$ \% & $ 0.9780$ \% &$0.0522$ \%& $ 0.7355$ \%  \\[1ex]
Case II ($w_{\nabla u}=0$) & $2.7517e-03 $&$1.4970$ \%& $2.8120$ \% &$0.4920$ \%& $ 2.2725$ \%   \\[1ex]
Case II ($w_{\nabla u}=1$) &$2.5277e-03 $&$0.1130 $ \%& $ 1.0973$ \%&$0.1630$ \%& $ 1.7915$ \% \\[1ex]
Case III ($w_{\nabla u}=0$) & $1.6881e-03$&$0.3170$ \% & $1.6801$ \%&$0.2430$ \% & $ 3.3520$ \%  \\[1ex]
Case III ($w_{\nabla u}=1$) &$1.4899e-03$&$0.7740$ \%& $ 1.2520$ \%&$0.0730 $ \%& $ 1.2185$ \%  \\[1ex]
Case IV ($w_{\nabla u}=0$) & $1.5556e-03  $&$0.5308$ \%& $1.1720$ \%&$0.1896$ \% & $ 1.2275$ \% \\[1ex]
Case IV ($w_{\nabla u}=1$)&$1.3695e-03$&$0.6666$ \%& $ 1.4820$ \%&$0.4230$ \% & $ 1.2555$ \%  \\[1ex]
\hline
\end{tabular}\label{tab:sig}
\end{center}
\end{table*}
%
Taken together, the results indicate that even with very few sensors very accurate
inference of the parameters as well as the field $u$ is obtained
using PINN. Moreover, penalizing the gradient of the measurements when possible
leads to better accuracy for $u$ although the improvement in the inference of
the parameters is mixed.

\subsection{Multi-fidelity data}
In many real-world applications, the training data is small and possibly inadequate to obtain even a rough estimation of the
parameters. Here, we demonstrate how we can resolve this issue by resorting to supplementary data of lower fidelity that may
come from cheaper instruments of lower resolution or from some computational models. We will refer to such data as ``low-fidelity"
and we will assume that we have a large number of such data points unlike the high-fidelity data.
Here, we will employ a composite network inspired by the recent work on multi-fidelity NNs in~\cite{Meng2019}.

The estimator of the high-fidelity model (HF) using the correlation structure to correct the low-fidelity model (LF), can be expressed as
\begin{align}
u_{HF}(x,t)=h(u_{LF}(x,t),x,t),
\end{align}
where $h$ is a correlation map to be learned, which is based on the correlation between the HF and LF data. Similarly,
we have two NN for low- and high-fidelity, respectively, as follows:
\begin{align}
u_{LF}&=\mathcal{NN}_{LF}(x_{LF},t_{LF},w_{LF},b_{LF}),\nonumber\\
u_{HF}&=\mathcal{NN}_{HF}(x_{HF},t_{HF},u_{LF},w_{HF},b_{HF}).\
\end{align}
We use 4 hidden layers and 20 neurons per layer for $\mathcal{NN}_{LF}$ and 2 hidden layers with 10 neurons for $\mathcal{NN}_{HL}$. The learning rate is $5*10^{-5}$.
We infer the same parameters as in the single-fidelity case and the field $u$ by minimizing the mean-squared-error loss function:
\begin{align}
MSE&=MSE_{u_{LF}}+MSE_{u_{HF}}+MSE_{f_{HF}},\
\end{align}
where
\begin{align}
MSE_{u_{LF}}&=\frac{1}{N_{LF}}\sum_{i=1}^{N_{LF}}|u_{LF}(t_{u_{LH}}^i,x_{u_{LF}}^i)-u_{LH}^i|^2,\nonumber\\
MSE_{u_{HF}}&=\frac{1}{N_{HF}}\sum_{i=1}^{N_{HF}}|u_{HF}(t_{u_{HF}}^i,x_{u_{HF}}^i)-u_{HF}^i|^2,\nonumber\\
MSE_{f_{HF}}&=\frac{1}{N_f}\sum_{i=1}^{N_f}|f_{HF}(t_{f_{HF}}^i,x_{f_{HF}}^i)|^2,\nonumber\
\end{align}
and $\{(t_{u_{LH}}^i,x_{u_{LH}}^i)\}_{i=1}^{N_{LF}}$ are the point of low-fidelity, $\{(t_{u_{HF}}^i,x_{u_{HF}})\}_{i=1}^{N_{HF}}$ are the point of high-fidelity, and $\{(t_{f_{HF}}^i,x_{f_{HF}}^i)\}_{i=1}^{N_f}$ are the residual points
where we penalize the residual $f$. We choose $N_f=1024$ for the tests here.

\begin{figure}[ht]
\begin{minipage}[]{0.5 \textwidth}
 \leftline{~~~~~~~\tiny\textbf{(a)}}
\centerline{\includegraphics[width=6cm,height=4cm]{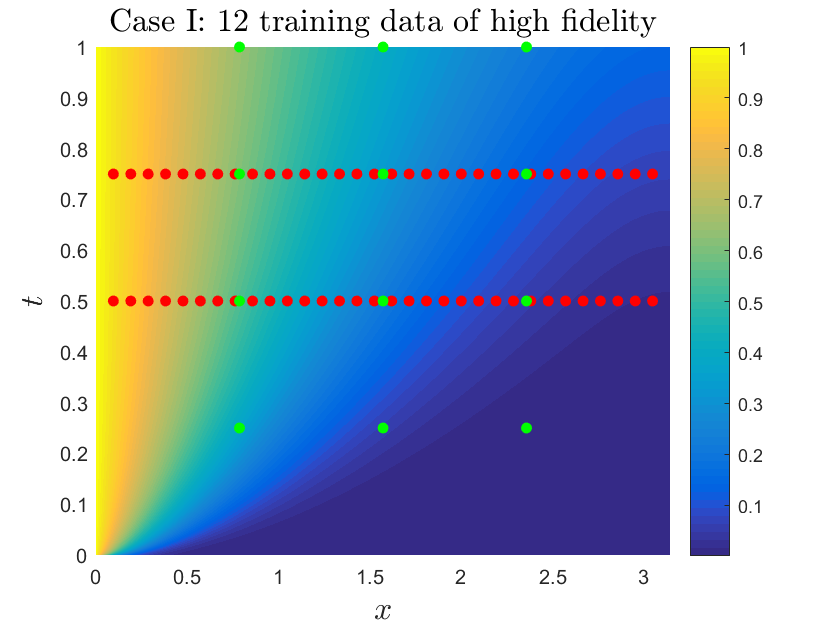}}
\end{minipage}
\begin{minipage}[]{0.5 \textwidth}
 \leftline{~~~~~~~\tiny\textbf{(b)}}
\centerline{\includegraphics[width=6cm,height=4cm]{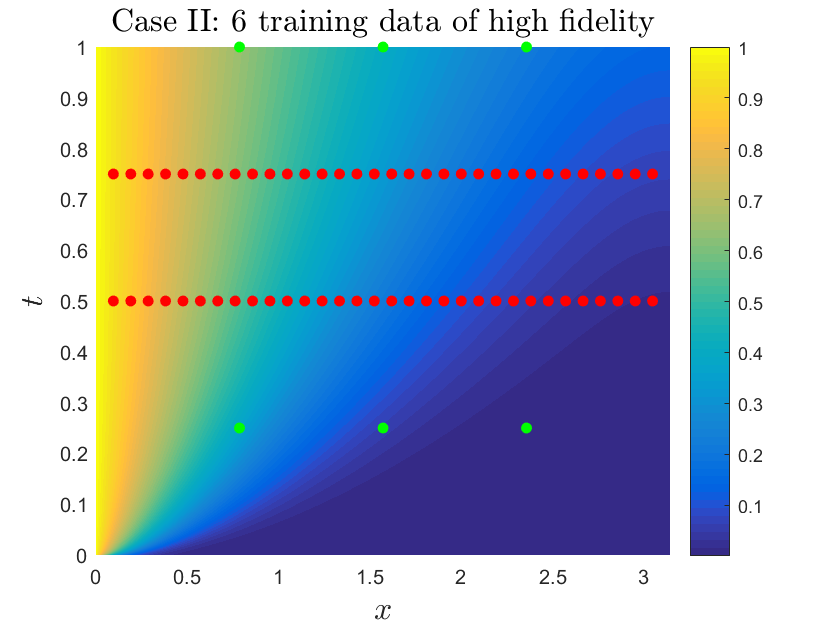}}
\end{minipage}
\caption{\textbf{} Multi-fidelity case: (a) 12 high-fidelity training data~($N_{HF}=12$). (b) 6 high-fidelity training data~~($N_{HF}=6$). }
\label{f3-2}
\end{figure}

We choose the reaction term $g(u)=\lambda_1 u^{\lambda_2}$, and set the true parameters $\nu_1=1$, $\nu_2=1$, $\lambda_1=-1$ and $\lambda_2=2$.
Here the low-fidelity training data is obtained by the second-order finite difference solution of \eqref{model}
with erroneous parameter values, i.e., $\nu_1=1.25$, $\nu_2=1.25$, $\lambda_1=-0.75$, and $\lambda_2=2.5$, where $\Delta x=\frac{\pi}{32}$ and $\Delta t=\frac{1}{32}$; we choose $64$ point of low-fidelity of $u$, i.e., $N_{LF}=64$. The positions of low-fidelity are denoted by the red point in Fig.~ \ref{f3-2}.
The high-fidelity data is obtained by the numerical solution of \eqref{model} when $\nu_1=1$, $\nu_2=1$, $\lambda_1=-1$, and $\lambda_2=2$ where $\Delta x=\frac{\pi}{1024}$ and $\Delta t=\frac{1}{1024}$. The positions of high-fidelity data are shown by the green points in Fig. \ref{f3-2}. We choose 12 high-fidelity training data ($N_{HF}=12$) in Fig. \ref{f3-2}(a) and 6 data as the high-fidelity training data ($N_{HF}=6$) in Fig. \ref{f3-2}(b).

To test the effect of the low-fidelity data, we compare the PINN and multi-fidelity PINN results in Table \ref{tab:mul}. As we can see, the parameter inference using the multi-fidelity PINN is much better than the single-fidelity predictions. Moreover, if we have a small number of HF data, e.g. $N_{HF}=6$, the results of the multi-fidelity PINN are still quite accurate.

\begin{table*}[ht]
\scriptsize
\begin{center}
\caption{Multi-fidelity case: Errors of the solution and of the parameters. (mPINN refers to
the multi-fidelity PINN)}
\begin{tabular}{ c cc cc cc cc cc c cc cc cc cc ccc c}
\hline
& $E_u$ &$E_{\nu_1} $ & $E_{\nu_2}$ &$E_{\lambda_1}$   & $E_{\lambda_2}$  \\[1ex]
\hline
12 point+PINN         & $2.3558e-03  $&$3.4000$ \%& $ 9.0891 $ \%&$1.6330   $ \% & $ 13.445$ \%   \\[1ex]
12 point+mPINN     &$1.1214e-03  $&$0.6380$ \% & $ 1.2772 $ \% &$2.4939     $ \%& $ 0.8975$ \% \\[1ex]
6 point+PINN          & $6.5386e-03  $&$9.3640 $ \%& $24.455 $ \%&$14.707$ \%& $ 49.445$ \% \\[1ex]
6 point+mPINN     &$1.2425e-03  $&$1.6190$ \%& $ 3.6775$ \%&$2.1220$ \%& $ 2.0635$ \%  \\[1ex]
\hline
\end{tabular}\label{tab:mul}
\end{center}
\end{table*}
\section{\label{Stoc}Results for the stochastic case}
Next, we test the effectiveness of sPINN for solving forward and inverse problems by considering the following
stochastic nonlinear ADR equation:
\begin{align}
\left\{
\begin{array}{ll}
u_t= (k(x;\omega)u_{x})_x\!-\!\nu_2 u_x\!+\!g(u)\!+\!f(x,t),
& (x,t,\omega) \in (x_0,x_1)\times(0,T]\times \Omega, \\
u(x,0)=1-x^2,
& x \in (x_0,x_1),\\
u(x_0,t)=0,~u(x_1,t)=0,
& t \in (0,T].
\end{array}\right.\label{model-1}
\end{align}
Here $x_0=0$, $x_1=1$, $\Omega$ is the random space, and
the stochastic diffusivity is modeled as $log(k(x;\omega))\in GP(k_0(x),Cov(x,x'))$, hence it is a non-Gaussian random process with mean $k_0(x)=\sin(\pi(x+1)/2)/5$, and covariance function $Cov(x,x')=\sigma^2*exp(-\frac{(x-x')^2}{l_c^2})$ with $l_c=1$ (GP stands for Gaussian Process here); $\sigma=0.1$.
We also define the residual $f=u_t-(k(x;\omega)u_{x})_x+\nu_2 u_x-g(u)$.
We consider the reaction term $g(u)=\lambda_1 u^{\lambda_2}$ and $f(x,t)=2$. The true parameter values are  $\nu_2=1$, $\lambda_1=1$, and $\lambda_2=3$.

\subsection{Forward problem}
We use a sPINN with 4 hidden layers and 20 neurons per layer for the modes of u$, i.e., u_l,~0\leq l \leq P$. The learning rate is $5*10^{-4}$.

We minimize the following mean-squared-error loss function:
\begin{align}
MSE&=MSE_{I}+MSE_{B}+MSE_{f},
\end{align}
where $MSE_{I}$ and $MSE_{B}$ are the loss functions for the initial and boundary conditions, respectively, and are computed as follows:
\begin{align}
MSE_{I}&=\frac{1}{N*N_u}\sum_{s=1}^{N}\sum_{i=1}^{N_I}[u_{NN}(x_u^{(i)},0;\omega_s)-u(x_u^{(i)},0;\omega_s)]^2,\nonumber\\
MSE_{B}&=\frac{1}{N*N_B}\sum_{s=1}^{N}\sum_{i=1}^{N_B}[u_{NN}(x_0,t_u^{(i)};\omega_s)-u(x_0,t_u^{(i)};\omega_s)]^2\nonumber\\
&~~~+\frac{1}{N*N_B}\sum_{s=1}^{N}\sum_{i=1}^{N_B}[u_{NN}(x_1,t_u^{(i)};\omega_s)-u(x_1,t_u^{(i)};\omega_s)]^2,\nonumber\\
MSE_{f}&=\frac{1}{N*N_f}\sum_{s=1}^{N}\sum_{i=1}^{N_f}[f_{NN}(x_f^{(i)},t_f^{(i)};\omega_s)-f(x_f^{(i)},t_f^{(i)};\omega_s)]^2.\nonumber\
\end{align}
We set $M=4$, $N=1000$, $N_I=101$, $N_B=101$ and $N_f=441$ in the loss function.

\begin{table*}[ht]
\scriptsize
\begin{center}
\caption{Forward problem: The $L_2$ error and the relative $L_2$ error for different values of the order $r$ of
arbitrary Polynomial Chaos (aPC).}
\begin{tabular}{ c cc cc cc cc cc c cc cc cc cc ccc c}
\hline
          &        & $L_2$  error      &    Relative $L_2$ error        \\[1ex]
\hline
$r=1$     &  $E[u]$& $ 2.5090e-03  $    & $3.0862e-03$ \\[1ex]
         &  $Var[u]$& $7.5067e-05 $  & $3.0841e-02$   \\[1ex]
\hline
$r=2$     &  $E[u]$& $ 1.0230e-03$    & $ 1.2583e-03 $\\[1ex]
         &  $Var[u]$& $4.4646e-06$ & $1.8343e-03 $\\[1ex]
\hline
$r=3$     &  $E[u]$& $6.1008e-04$    & $7.5045e-04$\\[1ex]
         &  $Var[u]$& $7.4720e-07$ & $3.0699e-04$ \\[1ex]
\hline
\end{tabular}\label{tab:1}
\end{center}
\end{table*}

\begin{figure}[htp]
\begin{minipage}[]{0.5 \textwidth}
\centerline{\includegraphics[width=6cm,height=4cm]{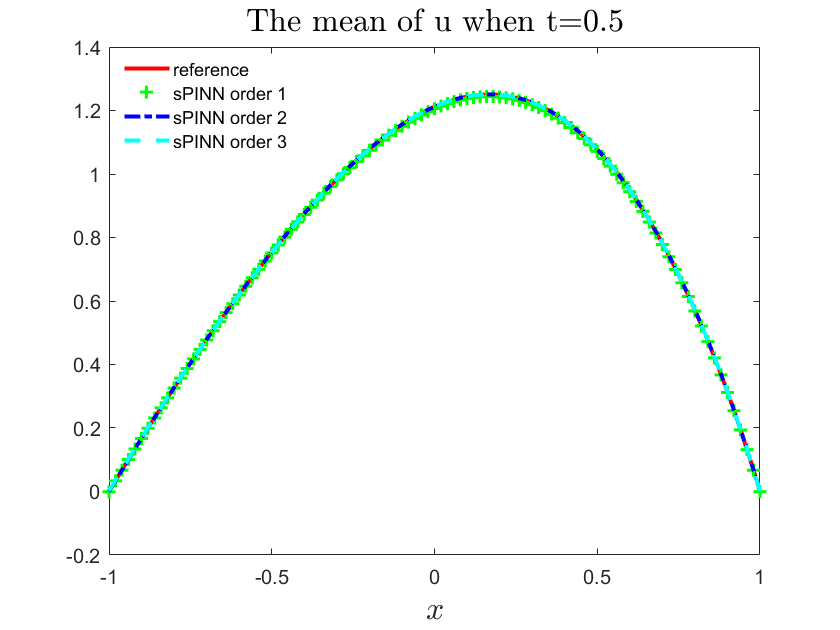}}
\end{minipage}
\hfill
\begin{minipage}[]{0.5 \textwidth}
\centerline{\includegraphics[width=6cm,height=4cm]{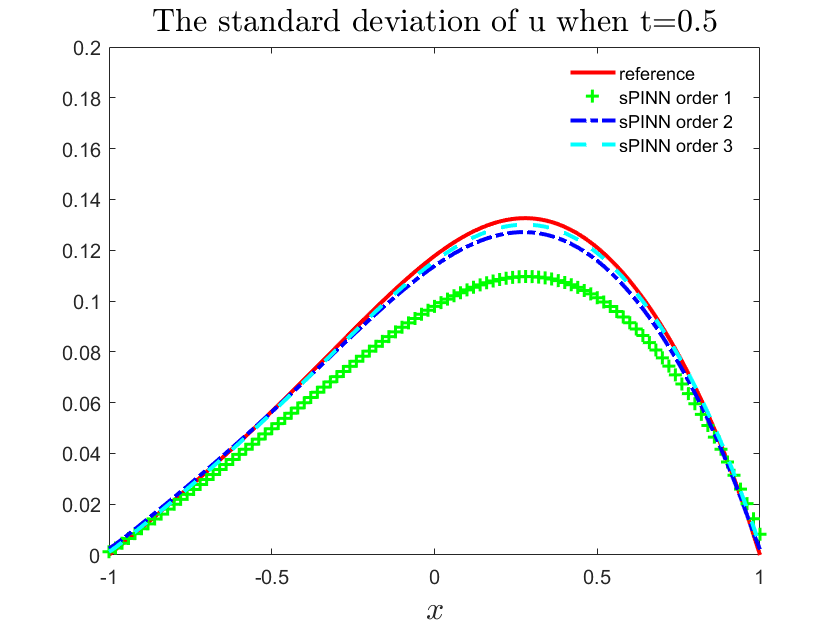}}
\end{minipage}
\caption{\textbf{} Forward problem: predicted mean and standard deviation at $t=0.5$ when $r=1,~2,~3$.
The reference solution is obtained by Quasi-Monte Carlo (see Appendix).}
\label{f3}
\end{figure}

We use the first-order ($r=1$, $P=4$), second-order ($r=2$, $P=14$) and third-order ($r=3$, $P=34$) aPC expansion for $u$ in this subsection. In Table \ref{tab:1} we present the $L_2$ and relative $L_2$ errors of the mean and variance of $u$ at $t=0.5$; using the higher order  aPC expansion, we obtain better results. We present in Fig. \ref{f3} the DNN predictions of the $u$ mean and variance at  $t=0.5$, where the reference solutions are calculated by the Qusi-Monte Carlo (QMC) method (more details are shown in Appendix). We also present in Fig. \ref{f4} the sPINN prediction of a few modes of $u$ at $t=0.5$. Taken together, the results show that sPINN can solve forward stochastic problems accurately for more complex (nonlinear and time-dependent)
 stochastic PDEs than the ones considered in the original paper of~\cite{zhang2019quantifying}.

\begin{figure}[htp]
\begin{minipage}[]{0.5\textwidth}
\centerline{\includegraphics[width=6cm,height=4cm]{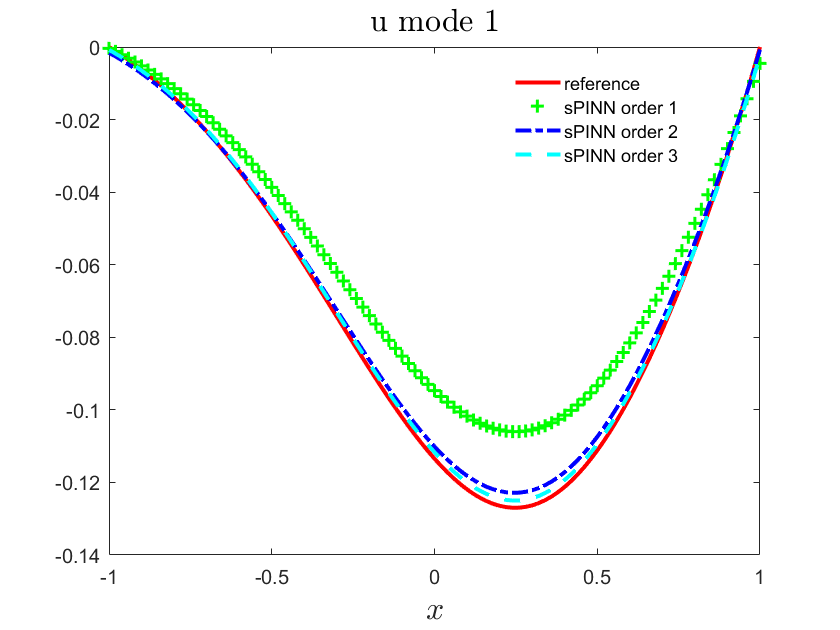}}
\end{minipage}
\begin{minipage}[]{0.5\textwidth}
\centerline{\includegraphics[width=6cm,height=4cm]{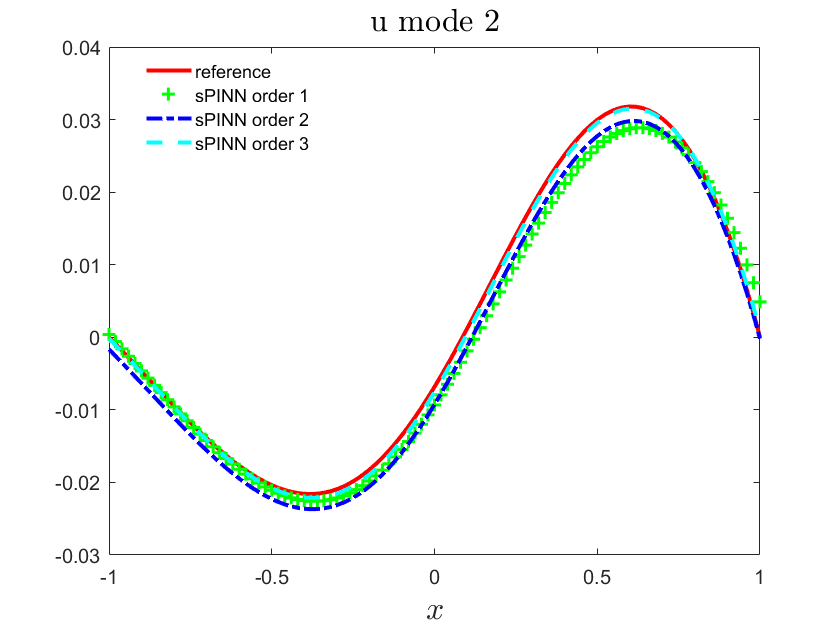}}
\end{minipage}
\begin{minipage}[]{0.5\textwidth}
\centerline{\includegraphics[width=6cm,height=4cm]{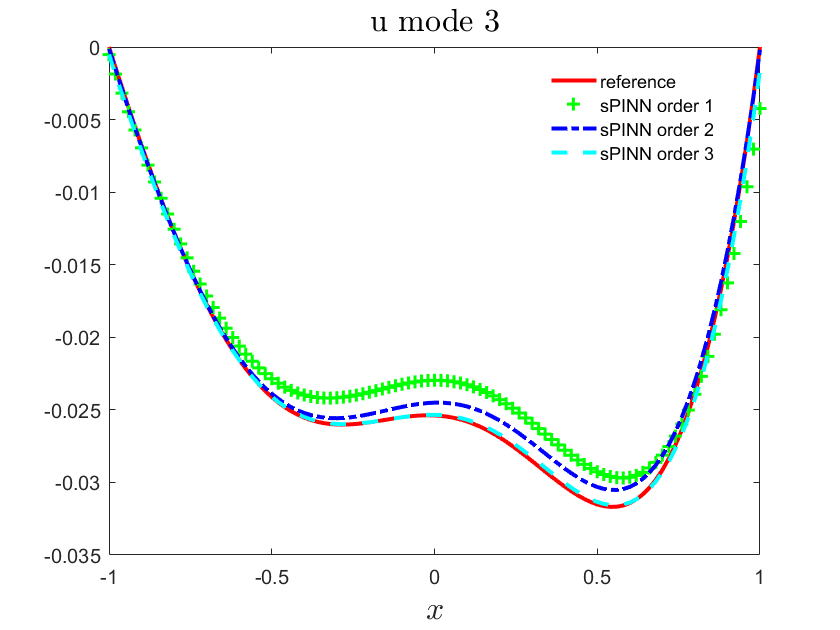}}
\end{minipage}
\hfill
\begin{minipage}[]{0.5\textwidth}
\centerline{\includegraphics[width=6cm,height=4cm]{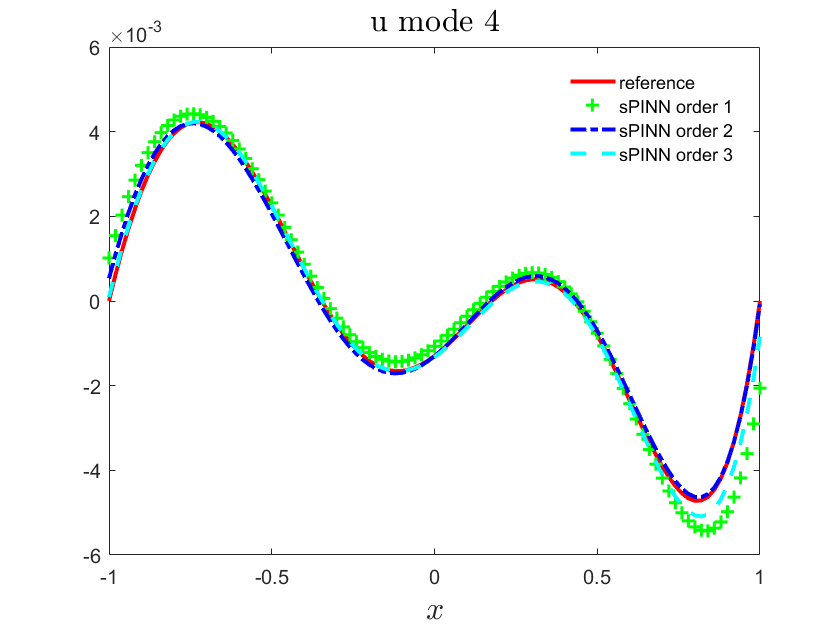}}
\end{minipage}
\hfill
\begin{minipage}[]{0.5\textwidth}
\centerline{\includegraphics[width=6cm,height=4cm]{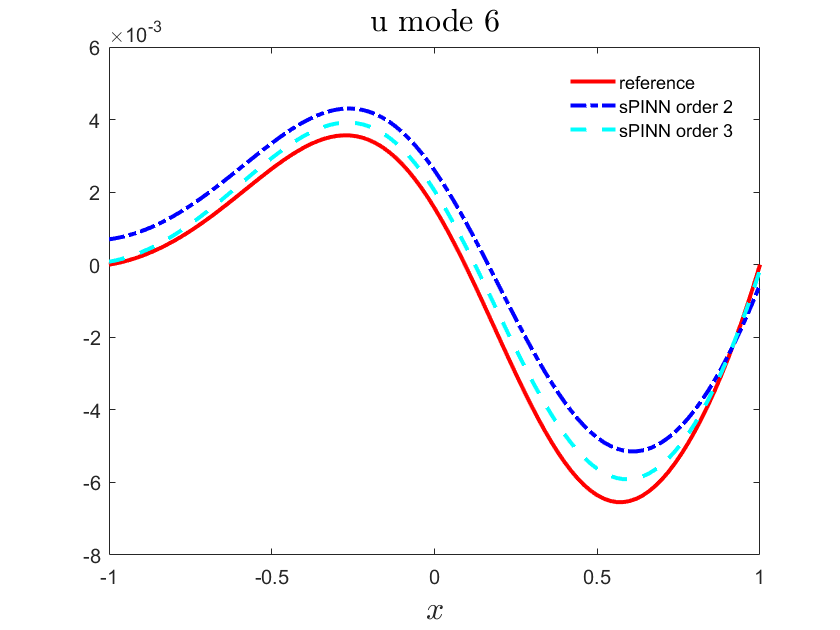}}
\end{minipage}
\hfill
\begin{minipage}[]{0.5\textwidth}
\centerline{\includegraphics[width=6cm,height=4cm]{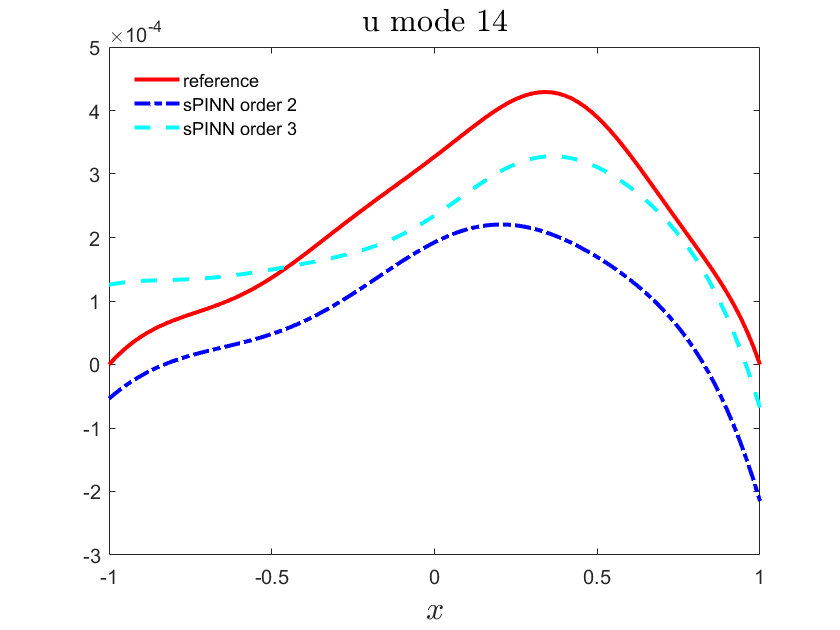}}
\end{minipage}
\hfill
\begin{minipage}[]{0.5\textwidth}
\centerline{\includegraphics[width=6cm,height=4cm]{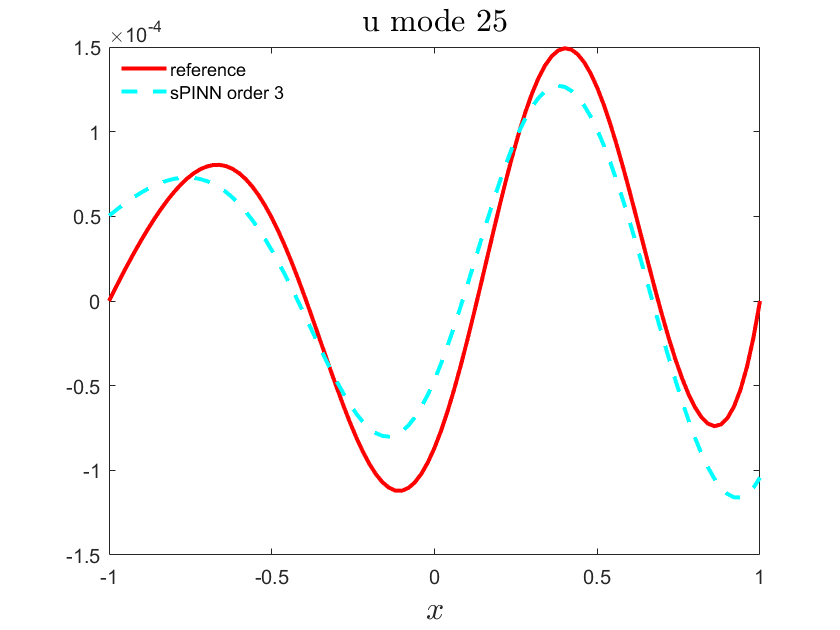}}
\end{minipage}
\hfill
\begin{minipage}[]{0.5\textwidth}
\centerline{\includegraphics[width=6cm,height=4cm]{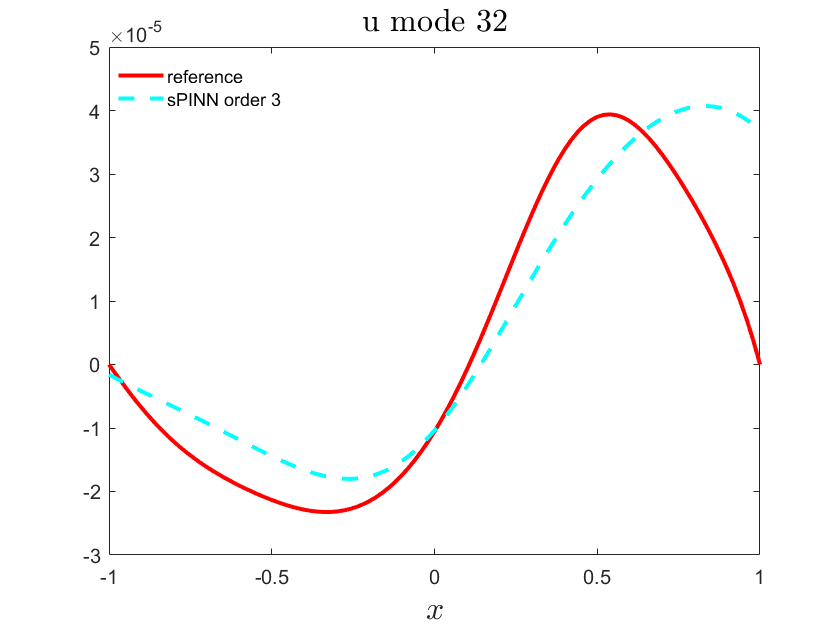}}
\end{minipage}
\caption{\textbf{} Forward problem: some predicted modes of $u$ with aPC expansions versus the reference solutions at $t=0.5$ for polynomial order $r=1,~2,~3$.}
\label{f4}
\end{figure}

\subsection{Inverse problem}
Next, we will infer the stochastic process $k(x,\omega)$ as well as the parameters $\nu_2$, $\lambda_1$, $\lambda_2$ and the solution $u(x,t,\omega)$.
%
We use 2 hidden layers and 4 neurons per layer for the $k$-mean and $k_i(x),~(i=1,...M)$ neural networks, and 4 hidden layers and 20 neurons per layer for the $u_{\alpha}(x,t),~(\alpha=0, 1, ...P)$ neural networks, and the learning rate is $5*10^{-4}$; we choose $N=2,000$, $N_u=20$, $N_k=7$, $N_f=441$, $w_u=100$ and $w_k=16$. These values of weights were chosen based on experimentation and also taken into account the order of
magnitude of the various quantities, e.g. mean versus standard deviation.

\begin{figure}[ht]
\centerline{\includegraphics[width=6cm,height=4cm]{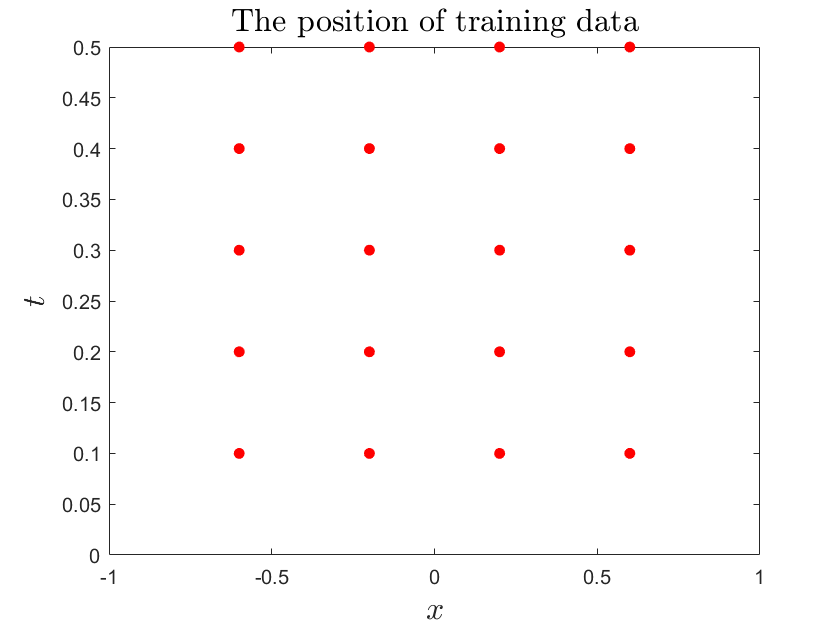}}
\caption{\textbf{} Stochastic inverse problem: Space-time positions of the training data for $u$. }
\label{f-sto-po}
\end{figure}

We minimize the following mean-squared-error loss function:
\begin{align}
MSE=10*(MSE_{u}+MSE_{k})+MSE_{f},
\end{align}
where
\begin{align}
MSE_{u}&=\frac{1}{N_u}\sum_{i=1}^{N_u}[\frac{1}{N}\sum_{s=1}^{N}u_{NN}(x_u^{(i)},t_u^{(i)};\omega_s)-\frac{1}{N}\sum_{s=1}^{N}u(x_u^{(i)},t_u^{(i)};\omega_s)]^2\nonumber\\
&~~~+w_u\frac{1}{N*N_u}\sum_{i=1}^{N_u}\sum_{s=1}^{N}[u_{NN}(x_u^{(i)},t_u^{(i)};\omega_s)-\frac{1}{N}\sum_{s=1}^{N}u_{NN}(x_u^{(i)},t_u^{(i)};\omega_s)\nonumber\\
&~~~+u(x_u^{(i)},t_u^{(i)};\omega_s)-\frac{1}{N}\sum_{s=1}^{N}u(x_u^{(i)},t_u^{(i)};\omega_s)]^2,\nonumber\\
MSE_{k}&=\frac{1}{N_k}\sum_{i\!=\!1}^{N_k}[\frac{1}{N}\sum_{s\!=\!1}^{N}k_{NN}(x_k^{(i)};\omega_s)\!-\!\frac{1}{N}\sum_{s\!=\!1}^{N}k(x_k^{(i)};\omega_s)]^2\!+\!w_k\frac{1}{N*N_u}\sum_{i\!=\!1}^{N_k}\sum_{s\!=\!1}^{N}\nonumber\\
&~~~[k_{NN}(x_k^{(i)};\omega_s)-\frac{1}{N}\sum_{s\!=\!1}^{N}k_{NN}(x_k^{(i)};\omega_s)+k(x_k^{(i)};\omega_s)\!-\!\frac{1}{N}\sum_{s\!=\!1}^{N}k(x_k^{(i)};\omega_s)]^2,\nonumber\\
MSE_{f}&=\frac{1}{N*N_f}\sum_{s\!=\!1}^{N}\sum_{i\!=\!1}^{N_f}[f_{NN}(x_f^{(i)},t_f^{(i)};\omega_s)-f(x_f^{(i)},t_f^{(i)};\omega_s)]^2.\nonumber\
\end{align}
We assume that we have measurements of $u$ at the positions indicated in Fig. \ref{f-sto-po};
the positions where $k$ is known are shown directly in the inference plots.

\begin{table*}[h]
\scriptsize
\begin{center}
\caption{$L_2$ and relative $L_2$ errors of $u$ and $k$. BO refers to the meta-learning results.}
\begin{tabular}{ c cc cc cc cc cc c cc cc cc cc ccc c}
\hline
  &                     & $E[u]$        & $Var[u] $   & $E[k]$     & $Var[k] $  \\[1ex]
\hline
$r=1$ &$L_2$ error&$2.0058e-03$&$6.0891e-07$&$2.0721e-03$&$1.1191e-06$\\[1ex]
&Relative $L_2$ error&$2.4673e-03 $  &$2.5017e-04$&$1.8106e-03$&$8.5212e-05$\\[1ex]
\hline
$r=2$&$L_2$  error  &$ 1.8472e-03$  &$8.2031e-07$ &$1.3080e-03$&$2.3888e-06$\\[1ex]
&Relative $L_2$ error&$2.2722e-03$&$3.3702e-04$&$1.1429e-03$ &$1.8189e-04$\\[1ex]
\hline
$r=3$&$L_2$  error   &$1.7582e-03$  &$5.9305e-07$ &$2.1860e-03$&$7.8451e-07$ \\[1ex]
&Relative $L_2$ error&$2.1628e-03$&$2.4365e-04$ &$1.9102e-03$ &$5.9735e-05$\\[1ex]
\hline
BO &$L_2$  error   &$5.9512e-04$  &$9.7195e-09$ &$3.9250e-04$&$1.5484e-07$\\[1ex]
&Relative $L_2$ error&$7.3204e-04$&$3.9933e-06$ &$3.4297e-04$ &$1.1790e-05$\\[1ex]
\hline
\end{tabular}\label{tab:2}
\end{center}
\end{table*}

\begin{table*}[h]
\scriptsize
\begin{center}
\caption{The error and the relative error of the parameters.}
\begin{tabular}{ c cc cc cc cc cc c cc cc cc cc ccc c}
\hline
  &                       & $E_{\nu_2} $  & $E_{\lambda1}$ &$E_{\lambda2} $ \\[1ex]
\hline
$r=1$    &  Error & $1.8855e-02$& $1.2252e-02 $    &$1.1636e-01 $ \\[1ex]
&Relative error   & $1.8855$ \% & $1.2252$ \%    &$3.8787$ \% \\[1ex]
\hline
$r=2$&  Error    & $2.3376e-02$& $4.2319e-04 $    &$6.3868e-02 $ \\[1ex]
       &Relative error& $2.3376$ \% &$0.0423$ \%    &$2.1289$ \%\\[1ex]
\hline
$r=3$& Error   & $1.9281e-02$& $2.306e-03 $    &$3.9592e-02 $ \\[1ex]
        &Relative error&$1.9281$ \% & $0.2306$ \% &$1.3197$ \% \\[1ex]
\hline
BO   & Error   & $2.2605e-03$& $9.2113e-04$&$2.1623e-02 $ \\[1ex]
    &Relative error& $0.2261$ \% & $0.0921$ \% &$0.7207$ \% \\[1ex]
\hline
\end{tabular}\label{tab:para}
\end{center}
\end{table*}
\newpage
We use the first-order ($r=1$ and $P=4$), second-order ($r=2$ and $P=14$) and third-order ($r=3$ and $P=34$) aPC expansions. The errors of the mean and variance of $u$ and $k$ are shown in Table \ref{tab:2}. The errors of the parameters are shown in Table \ref{tab:para}. Overall, the results improve by using a higher order aPC expansion.

\begin{figure}[htbp]
\begin{minipage}[]{0.5 \textwidth}
\centerline{\includegraphics[width=6cm,height=4cm]{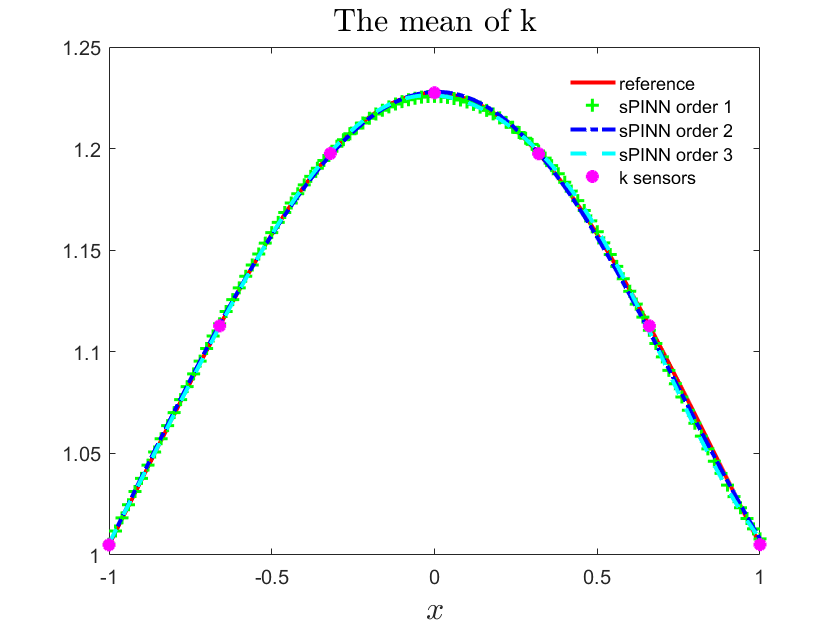}}
\end{minipage}
\hfill
\begin{minipage}[]{0.5 \textwidth}
\centerline{\includegraphics[width=6cm,height=4cm]{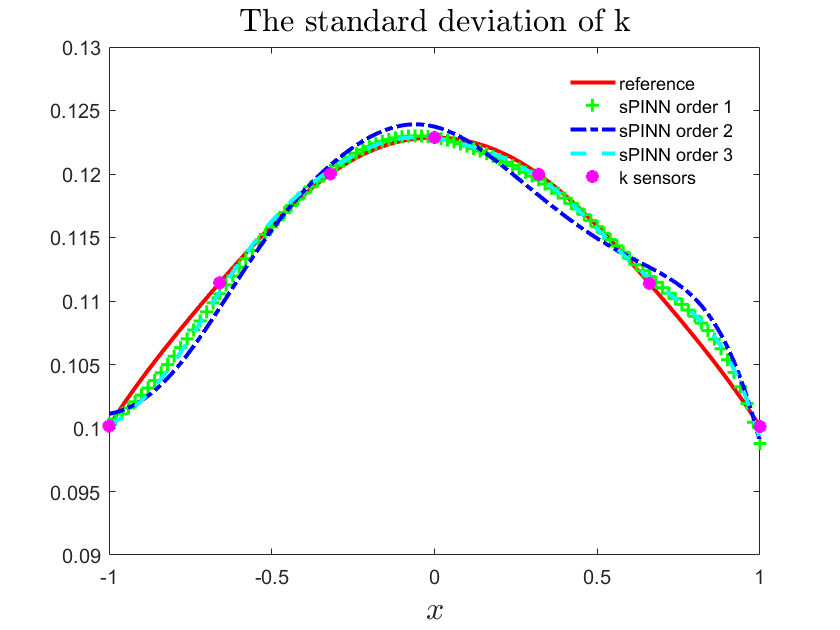}}
\end{minipage}
\begin{minipage}[]{0.5 \textwidth}
\centerline{\includegraphics[width=6cm,height=4cm]{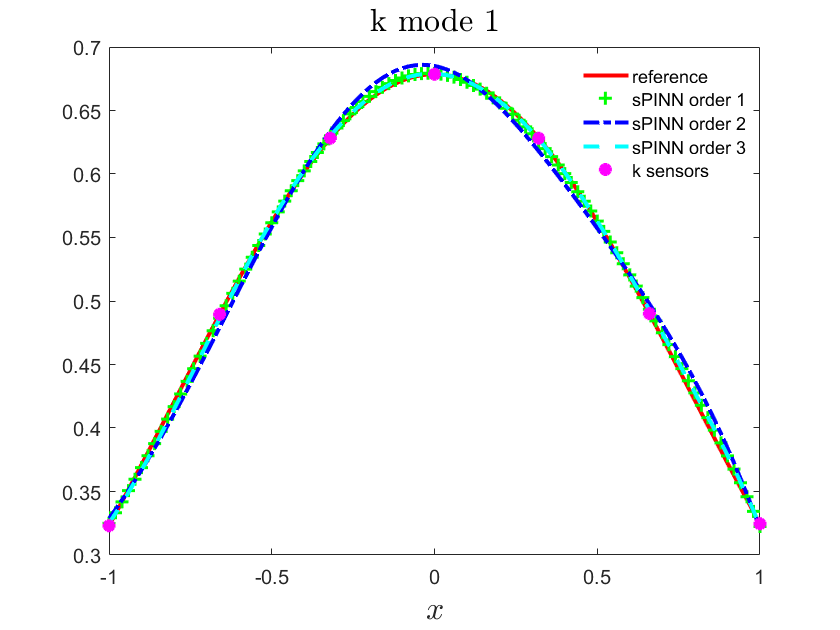}}
\end{minipage}
\hfill
\begin{minipage}[]{0.5 \textwidth}
\centerline{\includegraphics[width=6cm,height=4cm]{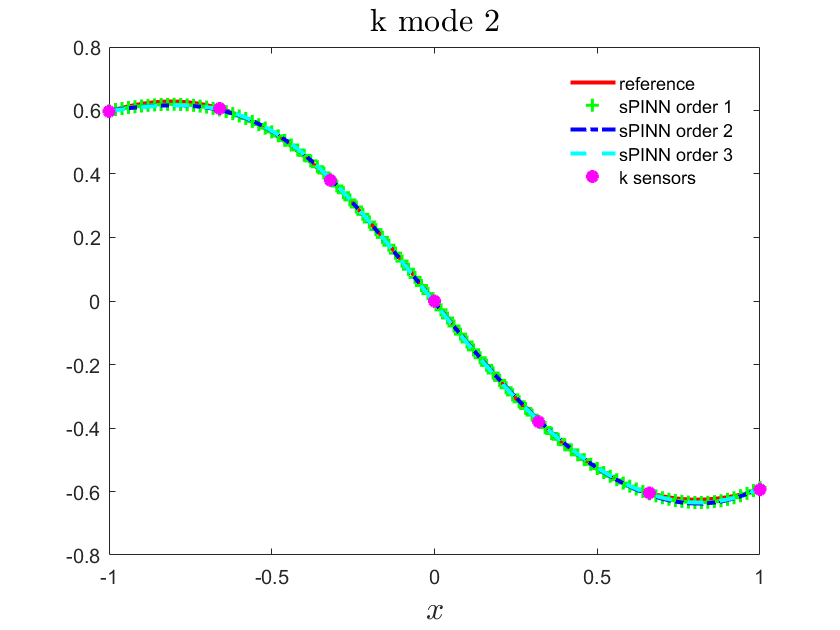}}
\end{minipage}
\hfill
\begin{minipage}[]{0.5 \textwidth}
\centerline{\includegraphics[width=6cm,height=4cm]{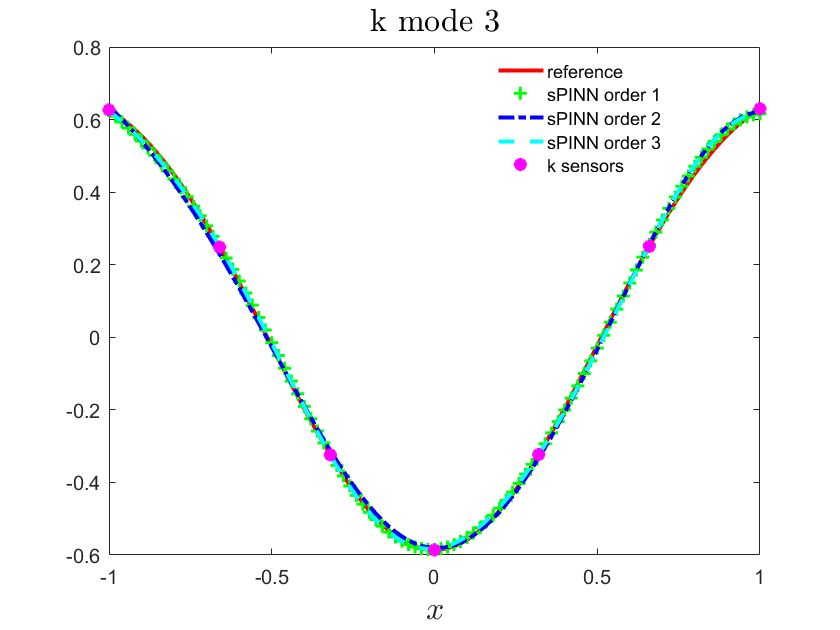}}
\end{minipage}
\hfill
\begin{minipage}[]{0.5 \textwidth}
\centerline{\includegraphics[width=6cm,height=4cm]{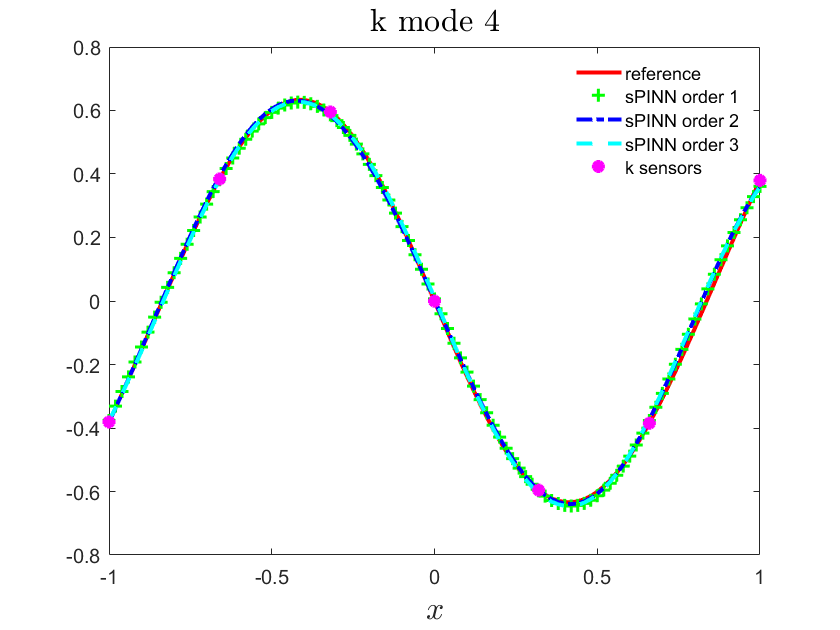}}
\end{minipage}
\caption{\textbf{} Stochastic inverse problem: predicted mean, standard deviation and modes of $k$ versus the reference solutions when $r=1,~2,~3$. The location of the k-sensors are denoted by red points.}
\label{f8}
\end{figure}

\begin{figure}[htbp]
\begin{minipage}[]{0.5 \textwidth}
\centerline{\includegraphics[width=6cm,height=4cm]{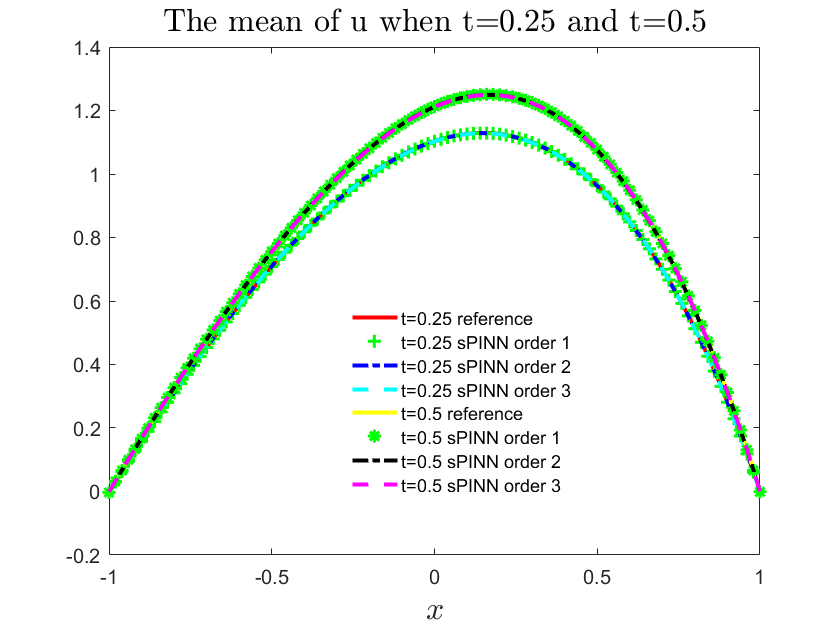}}
\end{minipage}
\hfill
\begin{minipage}[]{0.5 \textwidth}
\centerline{\includegraphics[width=6cm,height=4cm]{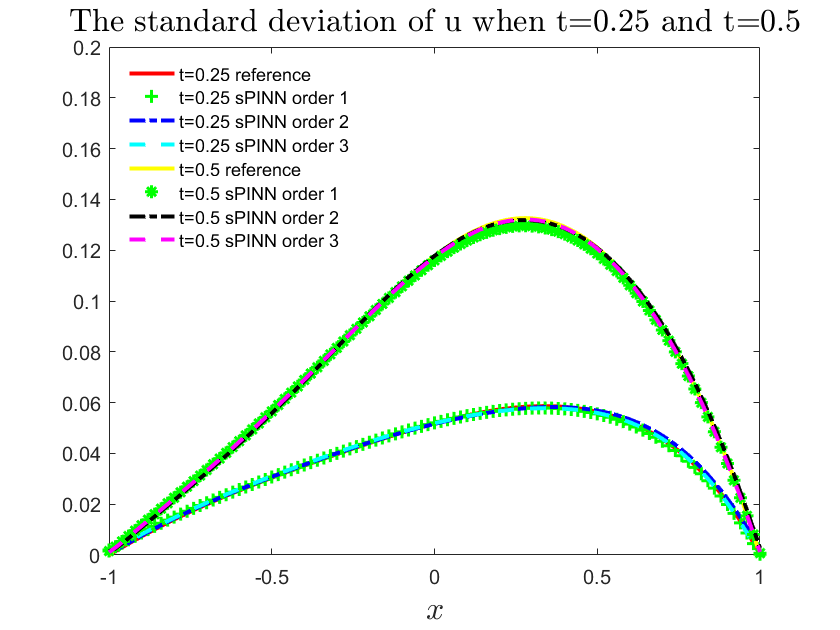}}
\end{minipage}
\begin{minipage}[]{0.5 \textwidth}
\centerline{\includegraphics[width=6cm,height=4cm]{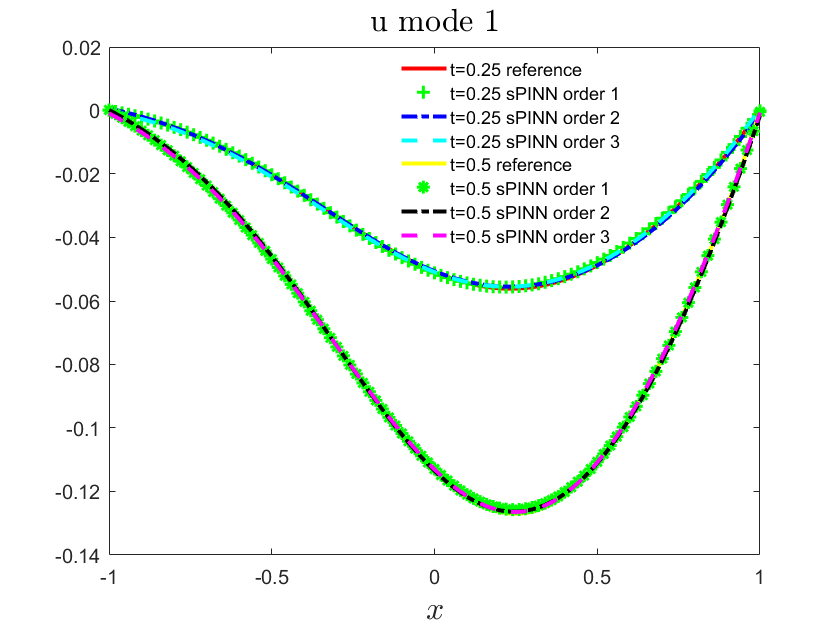}}
\end{minipage}
\hfill
\begin{minipage}[]{0.5 \textwidth}
\centerline{\includegraphics[width=6cm,height=4cm]{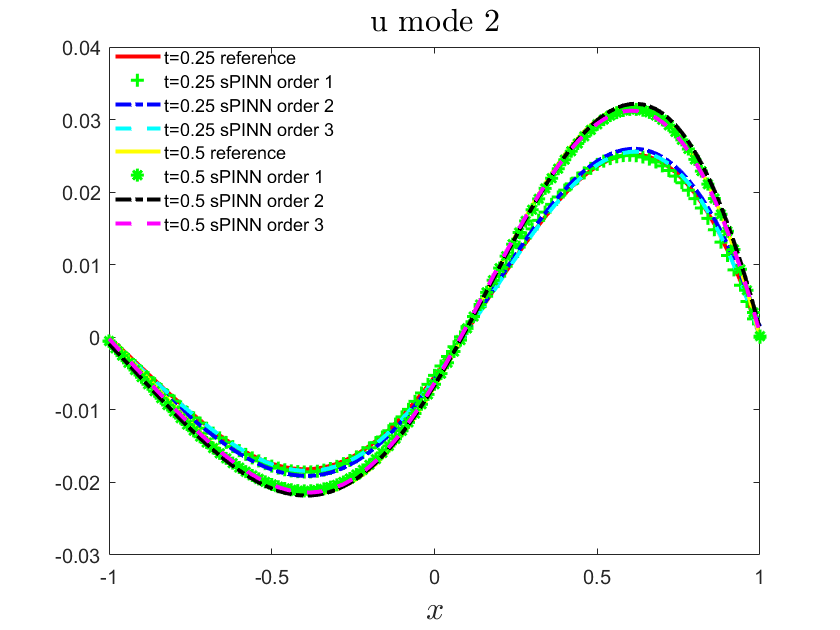}}
\end{minipage}
\hfill
\begin{minipage}[]{0.5 \textwidth}
\centerline{\includegraphics[width=6cm,height=4cm]{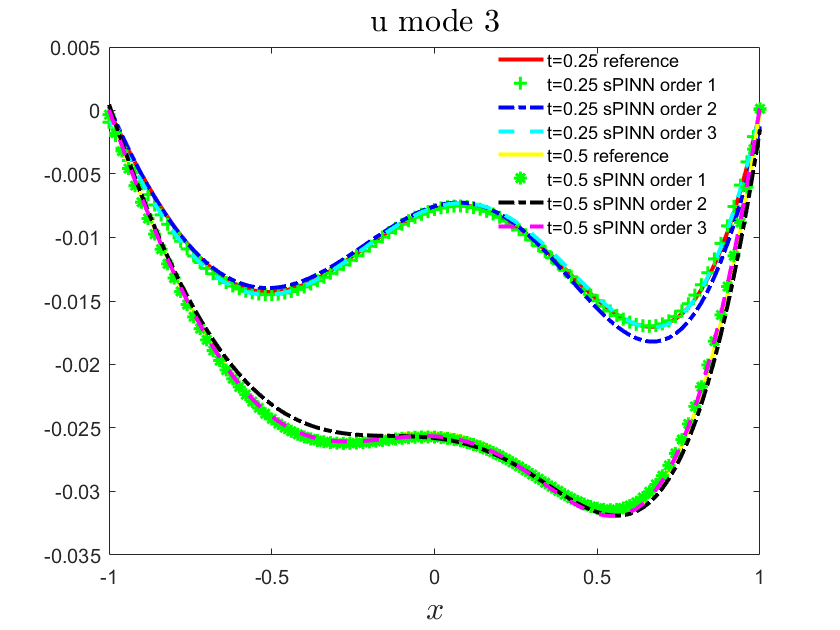}}
\end{minipage}
\hfill
\begin{minipage}[]{0.5 \textwidth}
\centerline{\includegraphics[width=6cm,height=4cm]{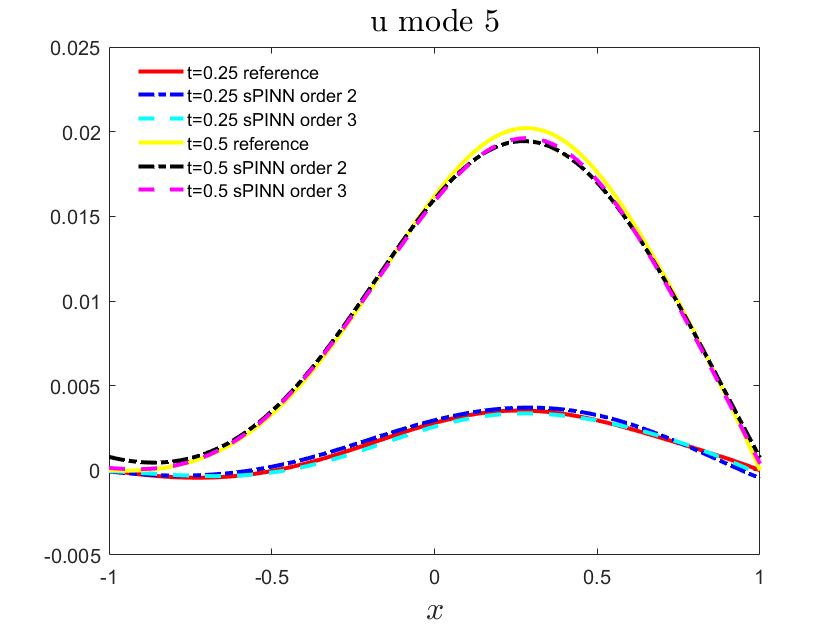}}
\end{minipage}
\begin{minipage}[]{0.5 \textwidth}
\centerline{\includegraphics[width=6cm,height=4cm]{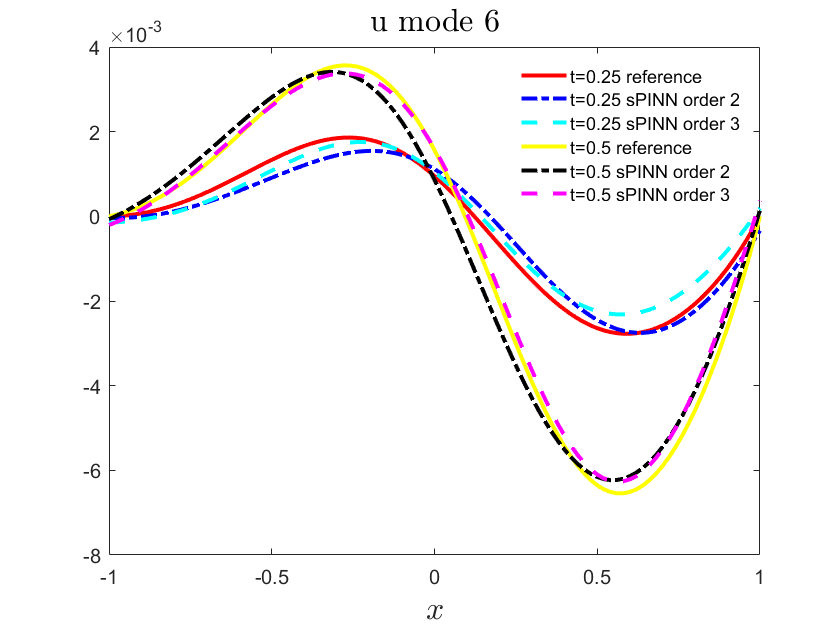}}
\end{minipage}
\hfill
\begin{minipage}[]{0.5 \textwidth}
\centerline{\includegraphics[width=6cm,height=4cm]{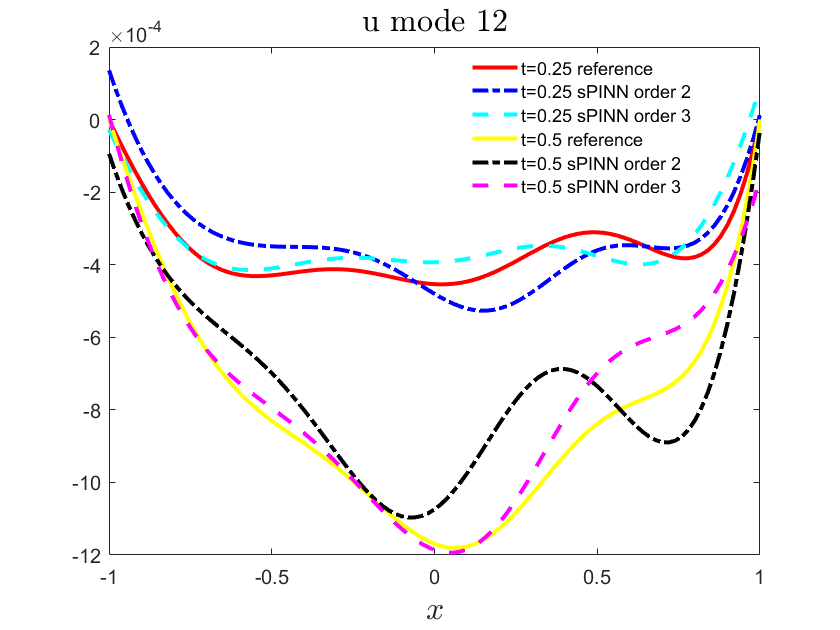}}
\end{minipage}
\caption{\textbf{} Stochastic inverse problem: predicted mean, standard deviation and modes of $u$ at $t=0.25$ and $t=0.5$ when $r=1,~2,~3$.}
\label{f10}
\end{figure}

The predicted mean, standard deviation and the modal functions of $k$ are shown in Fig. \ref{f8}. The predicted mean, standard deviation and the modal functions of $u$ are shown in Fig. \ref{f10}. The results for the solution $u$ are good but the inaccuracy of the first mode of $k$ affects the accuracy of the standard deviation. We have used a very small NN for $k$ as we observed problems with over-fitting, hence in order
to improve the overall learning of
$k$ in modal space, we will introduce the meta-learning method next to search for better NN architectures for $k$ but also for $u$.

\newpage
\subsection{Meta-learning}
To reduce the empiricism of selecting the sPINN architecture, in this section we employ Bayesian Optimization (BO) to learn the optimum structure of the NNs.
We use $d_K$ hidden layers and $w_K$ neurons per layer for $k$ mean neural network, and $d_U$ hidden layers and $w_U$  neurons per layer for $k_i(x),~(i=1,...M)$ and $u_{\alpha}(x,t),~(\alpha=0, 1, ...P)$ neural network. The learning rate is $l_r$.
The target is
\begin{align}
Target=10*(MSE_{u}+MSE_{k})+E_{\nu_2}+E_{\lambda_1}+E_{\lambda_2}.
\end{align}
So the target is a function: $\chi \rightarrow \mathbb{R}$, and $\chi=\{d_K, w_K,d_U,w_U, l_r  \}$.

Table \ref{tab:3} gives the range of the hyper-parameters we choose. We use the log-transform for the width of the NN and the learning rate. The top 10 good results are shown in Table \ref{tab:4}; the $*$ result denotes that we do not use the log-transform. These results suggest that we need a larger neural network for both $k$ and $u$.

\begin{table*}[ht]
\scriptsize
\begin{center}
\caption{The hyper-parameters and architecture choices for the fully
connected neural networks.}
\begin{tabular}{ c cc cc cc cc cc c cc cc cc cc ccc c}
\hline
&Hyper-parameter           & Range   & Log-transform        \\[1ex]
& hidden layers ($d_K$)      &$[1,10]$     &No         \\[1ex]
& units per layer ($w_K$)   &$[1,64]$   &Yes        \\[1ex]
& hidden layers  ($d_U$)     &$[1,30]$    &No        \\[1ex]
& units per layer ($w_U$)   &$[1,128]$     &Yes         \\[1ex]
&learning rate ($log(l_r)$)&$[-5,-2]$    &Yes        \\[1ex]
\hline
\end{tabular}\label{tab:3}
\end{center}
\end{table*}

\begin{table*}[ht]
\scriptsize
\begin{center}
\caption{Top 10 results of meta-learning using Bayesian optimization.}
\begin{tabular}{ c cc cc cc cc cc c cc cc cc cc ccc c}
\hline
&Number& Target  & $d_K$   & $w_K$   &  $d_U$  &  $w_U$  & $log(l_r)$       \\[1ex]
&1       &0.03898   &3       &23       & 10   &  113    & $-3.5670$ \\[1ex]
&2       &0.04678    &3       &21       & 9  &  107    & $-2.3724$ \\[1ex]
&3       &0.04704    &3       &23       &  10    &  113    & $-3.6665$ \\[1ex]
&4       &0.04915    &2       &2       &  6    &  7    & $-2.3740$ \\[1ex]
&5       &0.04962    &3       &21       &  9    &  127    & $-3.4890$ \\[1ex]
&6       &0.05007    &3       &25      &  9    &  118    & $-3.7266$ \\[1ex]
&7       &0.05429    &3       &22       & 9   & 110    & $-3.6487$ \\[1ex]
&8       &0.05451   &3       &22       & 9    & 118    & $-3.6606$ \\[1ex]
&9       &0.05691    &3       &20       & 9   &  107    & $-3.7656$ \\[1ex]
&10       &0.05735   &2       &2       &  6    &  7    & $-3.1180$ \\[1ex]
&*       &0.04704   &3       &4       &  15    &  80    & $-3.4890$  \\[1ex]
\hline
\end{tabular}\label{tab:4}
\end{center}
\end{table*}

\begin{figure}[htbp]
\begin{minipage}[]{0.5 \textwidth}
\centerline{\includegraphics[width=6cm,height=4cm]{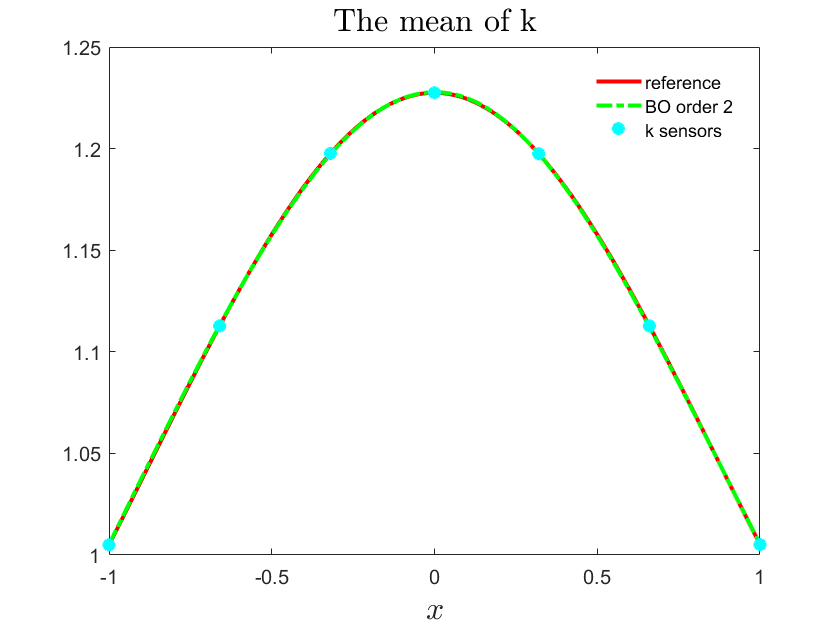}}
\end{minipage}
\hfill
\begin{minipage}[]{0.5 \textwidth}
\centerline{\includegraphics[width=6cm,height=4cm]{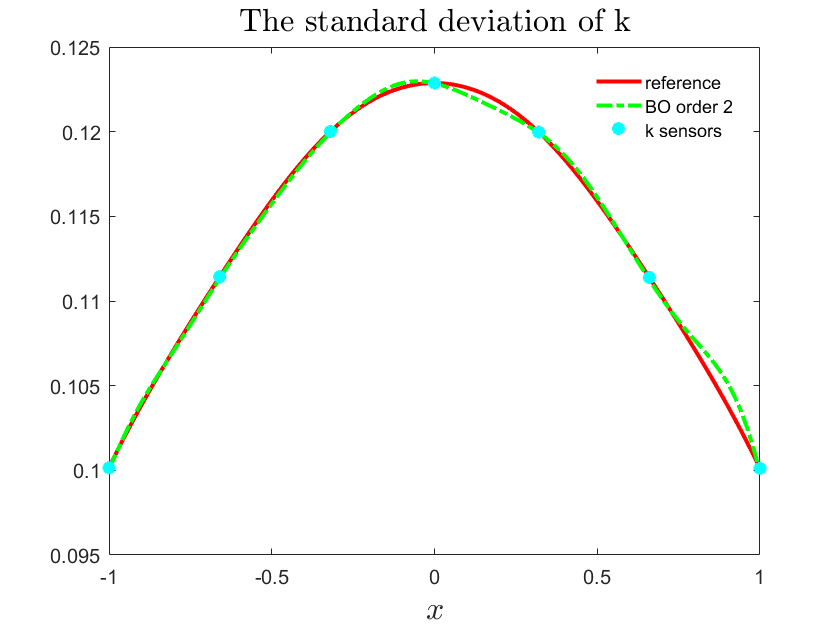}}
\end{minipage}
\begin{minipage}[]{0.5 \textwidth}
\centerline{\includegraphics[width=6cm,height=4cm]{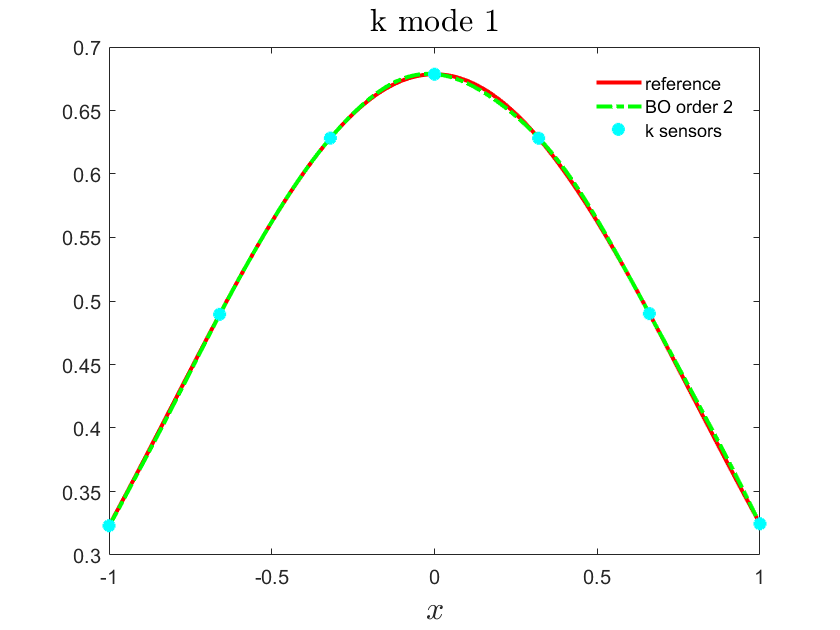}}
\end{minipage}
\hfill
\begin{minipage}[]{0.5 \textwidth}
\centerline{\includegraphics[width=6cm,height=4cm]{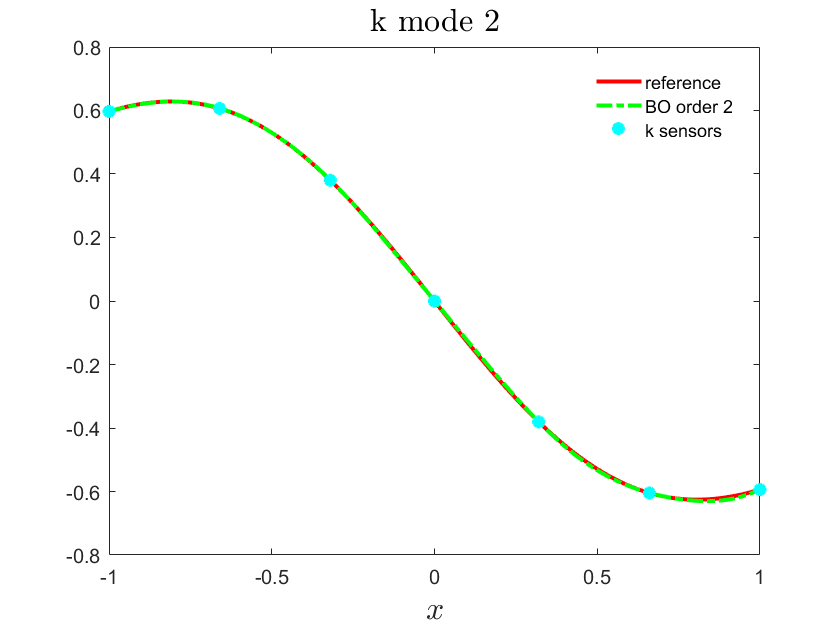}}
\end{minipage}
\hfill
\begin{minipage}[]{0.5 \textwidth}
\centerline{\includegraphics[width=6cm,height=4cm]{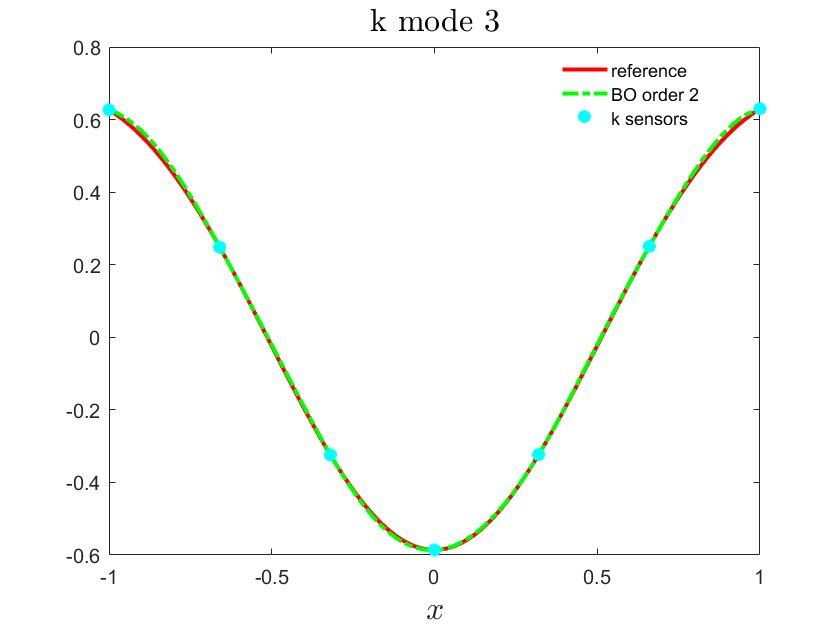}}
\end{minipage}
\hfill
\begin{minipage}[]{0.5 \textwidth}
\centerline{\includegraphics[width=6cm,height=4cm]{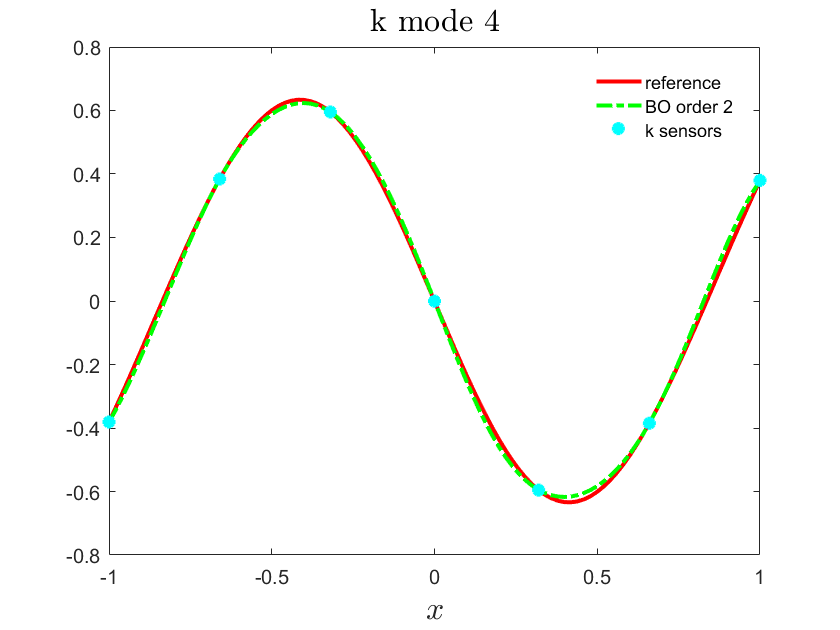}}
\end{minipage}
\caption{\textbf{} Stochastic inverse problem: Predicted results of sPINN for $k$ against the reference solution
using the optimum hyper-parameters obtained via
Bayesian Optimization (meta-learning).}
\label{f-BO-KMOD}
\end{figure}

\begin{figure}[htbp]
\begin{minipage}[]{0.5 \textwidth}
\centerline{\includegraphics[width=6cm,height=4cm]{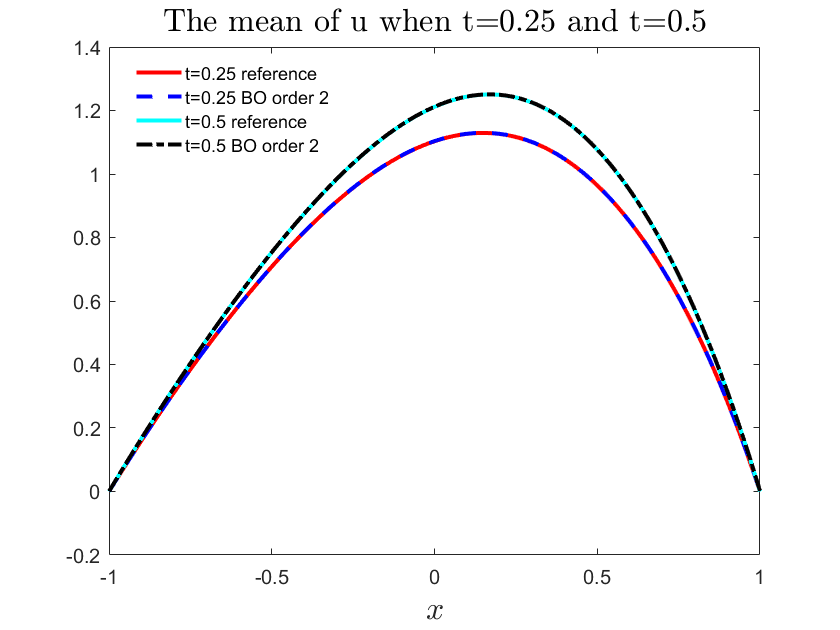}}
\end{minipage}
\begin{minipage}[]{0.5 \textwidth}
\centerline{\includegraphics[width=6cm,height=4cm]{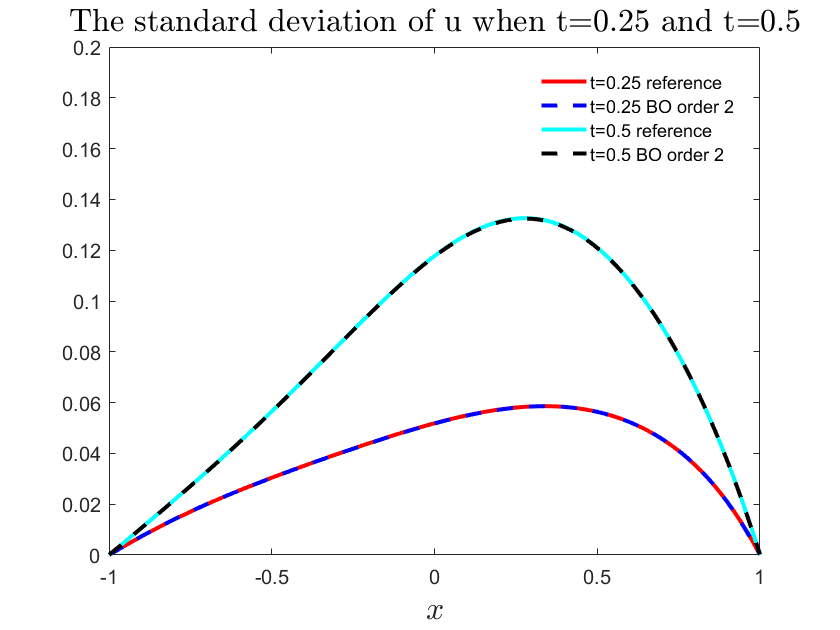}}
\end{minipage}
\begin{minipage}[]{0.5 \textwidth}
\centerline{\includegraphics[width=6cm,height=4cm]{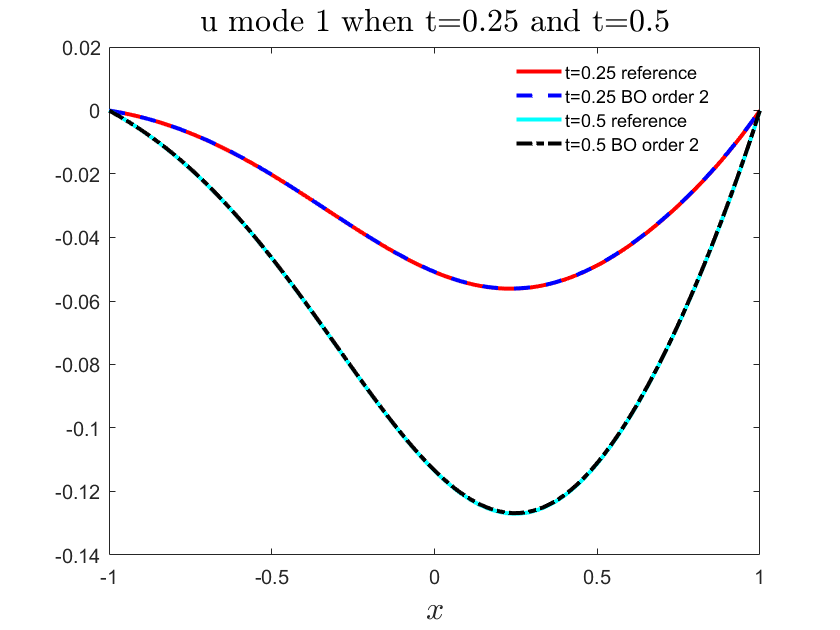}}
\end{minipage}
\hfill
\begin{minipage}[]{0.5 \textwidth}
\centerline{\includegraphics[width=6cm,height=4cm]{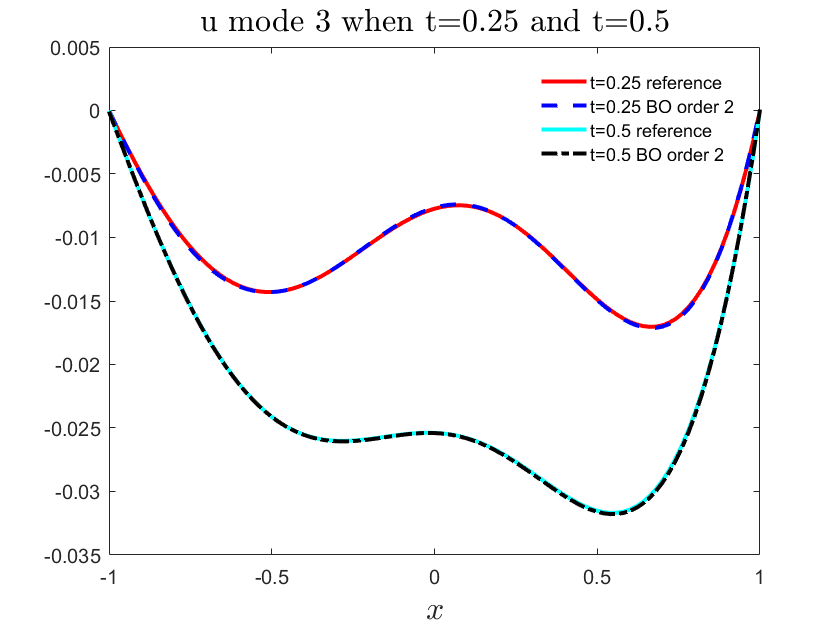}}
\end{minipage}
\hfill
\begin{minipage}[]{0.5 \textwidth}
\centerline{\includegraphics[width=6cm,height=4cm]{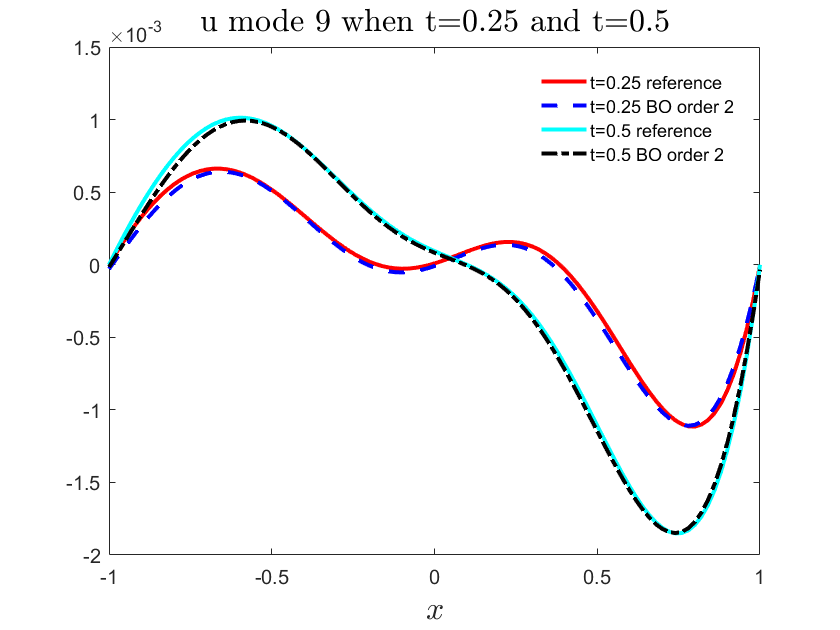}}
\end{minipage}
\hfill
\begin{minipage}[]{0.5 \textwidth}
\centerline{\includegraphics[width=6cm,height=4cm]{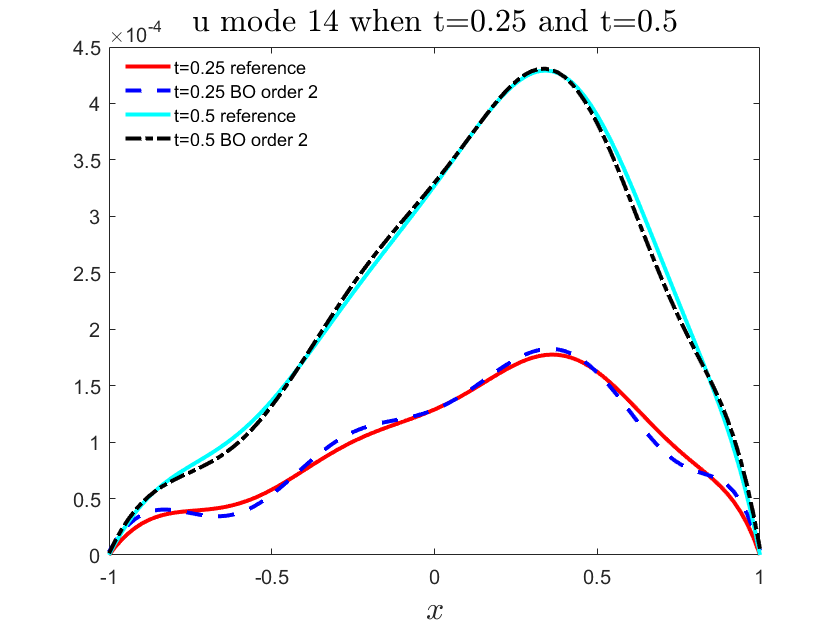}}
\end{minipage}
\caption{\textbf{} Stochastic inverse problem: Predicted results of sPINN for $u$
against the reference solutions at $t=0.25$ and $t=0.5$ using the optimum hyper-parameters obtained via
Bayesian Optimization (meta-learning).}
\label{f-BO-UMOD}
\end{figure}

Next, we use the best structure of the neural network and learning rate to re-compute the previous stochastic inverse problem, i.e. the depth of $k$ mean neural network is $3$, and the width is $23$. For the neural network of the $k$ modal functions, the $u$ mean and the $u$ modal functions, the depth is $10$ and the width is $113$. The learning rate is $10^{-3.5670}$. The results are shown in Figs. \ref{f-BO-KMOD}-\ref{f-BO-UMOD}.

\begin{figure}[htbp]
\centerline{\includegraphics[width=9cm,height=6cm]{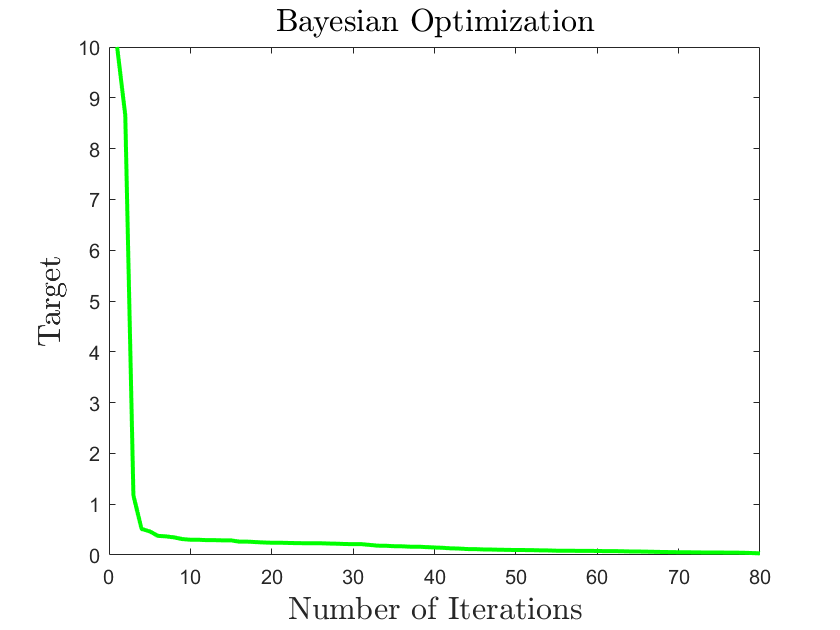}}
\caption{\textbf{} Convergence of the target for the Bayesian Optimization (meta-learning)
as a function of the number of iteration.}
\label{f-BO}
\end{figure}

\section{Summary}
We addressed here a special inverse problem governed by a stochastic nonlinear advection-diffusion-reaction (ADR) equation,
where given some samples of the solution $u(x,t;\omega)$ at a relatively few locations (here 4 spatial locations and 5 time instants)
but also given some samples of the stochastic diffusivity  $k(x;\omega)$ at 7 locations, we aim to obtain the full stochastic fields
for $u$ and $k$ as well as 3 other unknown parameters. We designed composite neural networks (NNs), including NNs induced by the
stochastic ADR equation, and relied on spectral expansions to represent stochasticity in order to deal with the sparsity in data.
We also presented a Bayesian Optimization method for learning the hyper-parameters of this composite NN as it is time consuming
to find the proper NNs by trial and error. We followed a hierarchical approach in testing the various components of the NNs,
including training from multi-fidelity data, investigating possible good locations in space-time for collecting the training
data, and evaluating different weights in the loss functions for the multiple terms representing data and physics. To the best
of our knowledge, this is the first time that such a study is undertaken with the purpose of evaluating the potential of
NNs to learn from sparse data of variable-fidelity and with uncertainty.

An important component missing in our study is
quantifying the uncertainty of the NN approximation as was first done in related work in~\cite{zhang2019quantifying} addressing
the {\em total} uncertainty. This is a serious but complex issue requiring the use of multiple methods to interpret this uncertainty
in an objective way, and we will pursue this line of research in future work. The present work is also the first study that uses
meta-learning for PINNs, i.e., to
optimize the composite NN, which in our case consists of multiple NNs, as would be the case in simulating multi-physics dynamics.
In addition to the Bayesian optimization employed here, one could also consider using several other methods, including genetic algorithms \cite{genetic-mitchell},
the greedy method \cite{li2009coordinate}, hyperband \cite{li2016hyperband,falkner2018bohb} as well as blended versions of the aforementioned methods or even another NN, like an RNN in conjunction
with reinforcement learning~\cite{EAS}, to search for the best architecture. This has already been done for classification work and it is part of AutoML \cite{he2018amc} but not for regression tasks.

\section*{Acknowledgement}
This work was supported by China Scholarship Council scholarship and the PhILMs grant DE-SC0019453. In addition, we would like to thank Dr. Guofei Pang, Dr.
Lu Lu, Dr. Xuhui Meng and Dr. Dongkun Zhang in the Division of Applied Mathematics at Brown University for their helpful suggestions.

\section*{Appendix}
\begin{figure}[htbp]
\begin{minipage}[]{0.5 \textwidth}
\centerline{\includegraphics[width=6cm,height=4cm]{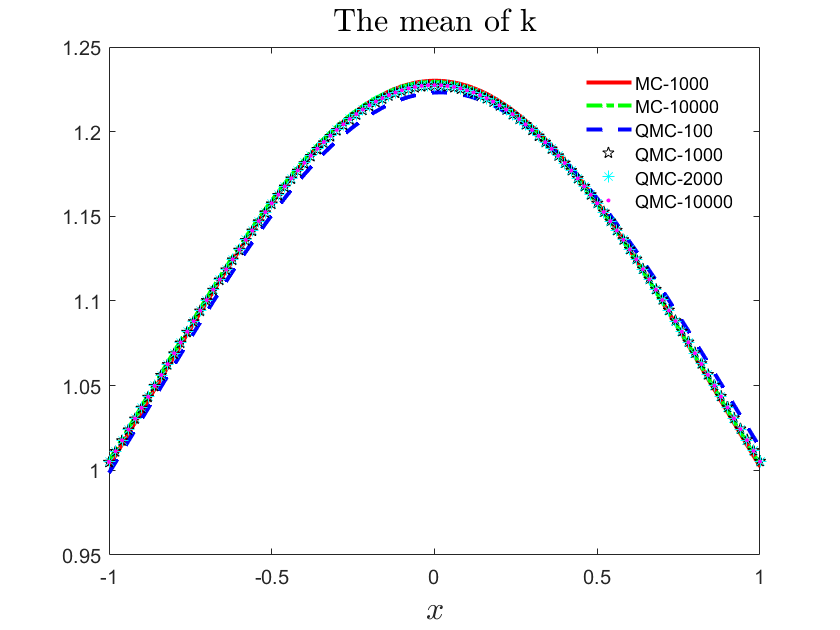}}
\end{minipage}
\hfill
\begin{minipage}[]{0.5 \textwidth}
\centerline{\includegraphics[width=6cm,height=4cm]{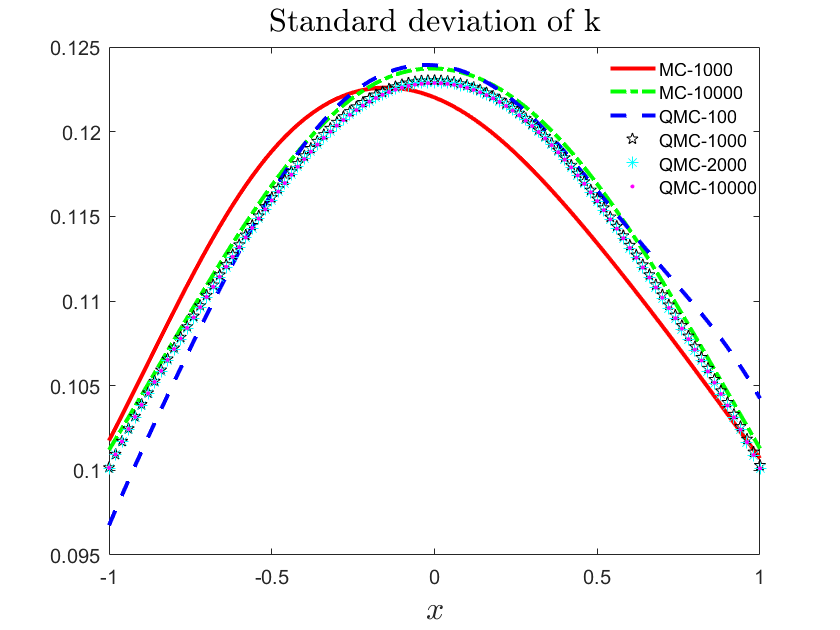}}
\end{minipage}
\begin{minipage}[]{0.5 \textwidth}
\centerline{\includegraphics[width=6cm,height=4cm]{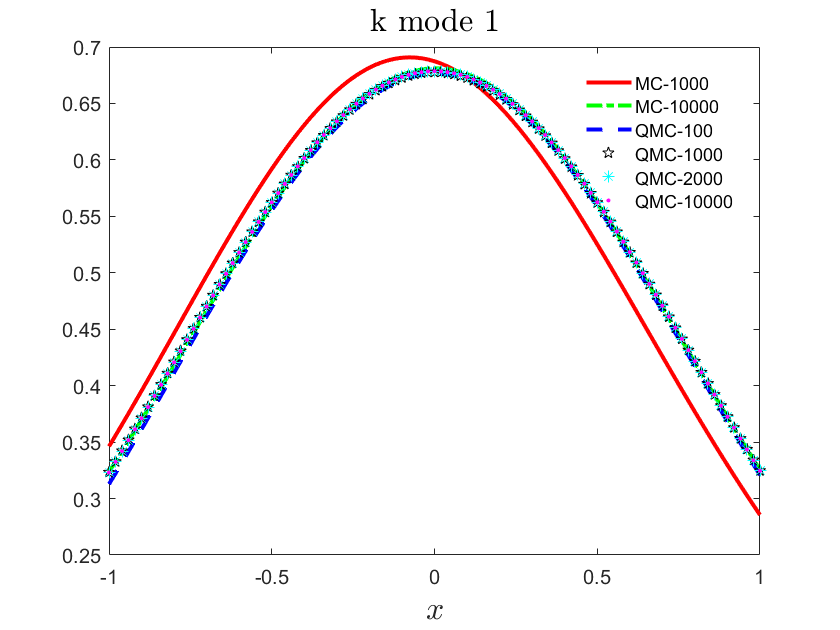}}
\end{minipage}
\hfill
\begin{minipage}[]{0.5 \textwidth}
\centerline{\includegraphics[width=6cm,height=4cm]{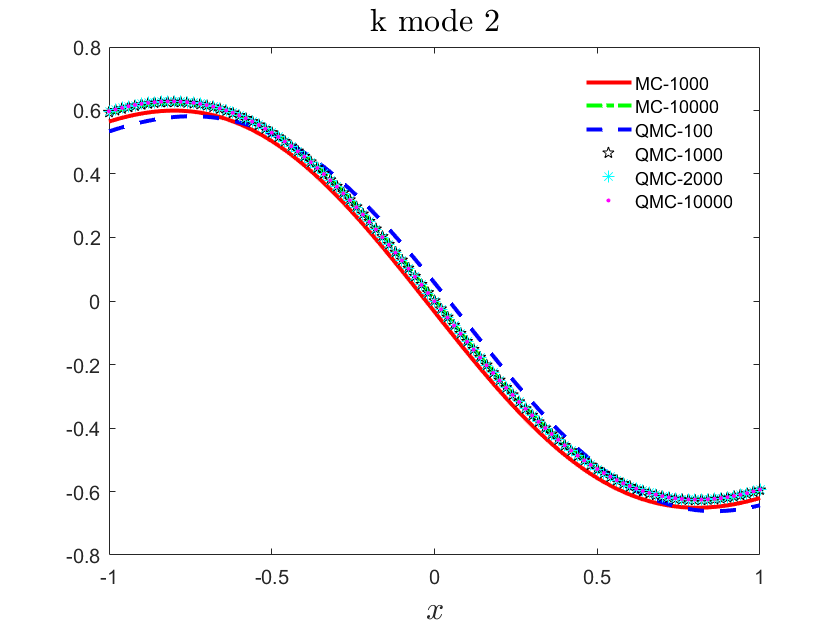}}
\end{minipage}
\hfill
\begin{minipage}[]{0.5 \textwidth}
\centerline{\includegraphics[width=6cm,height=4cm]{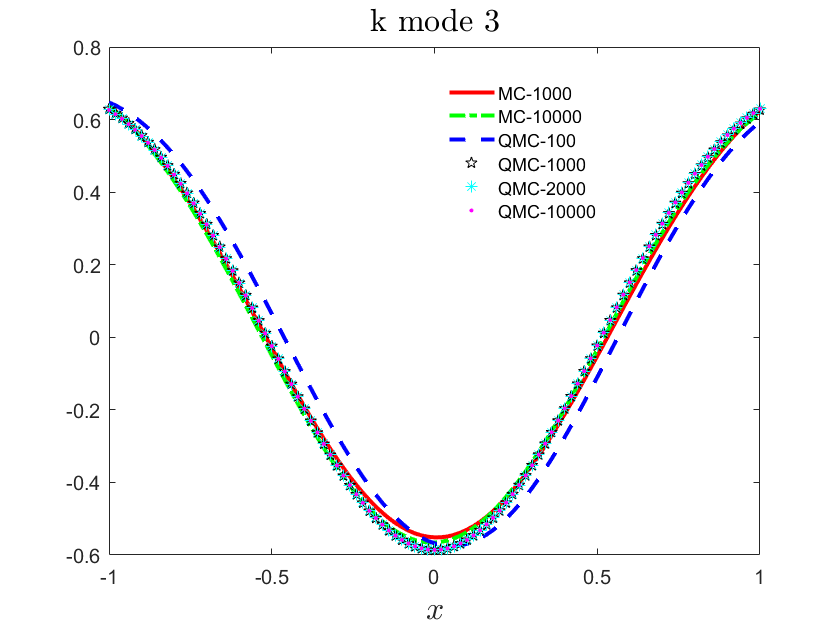}}
\end{minipage}
\hfill
\begin{minipage}[]{0.5 \textwidth}
\centerline{\includegraphics[width=6cm,height=4cm]{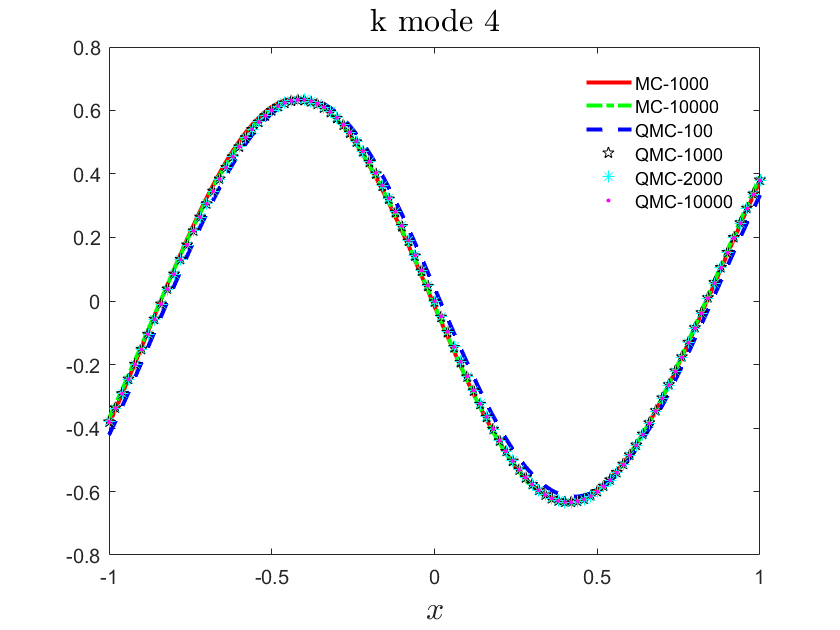}}
\end{minipage}
\caption{\textbf{} The mean, standard deviation and mode functions of k: MC vs QMC.}
\label{fQMC-K1}
\end{figure}

\begin{figure}[htbp]
\begin{minipage}[]{0.5 \textwidth}
\centerline{\includegraphics[width=6cm,height=4cm]{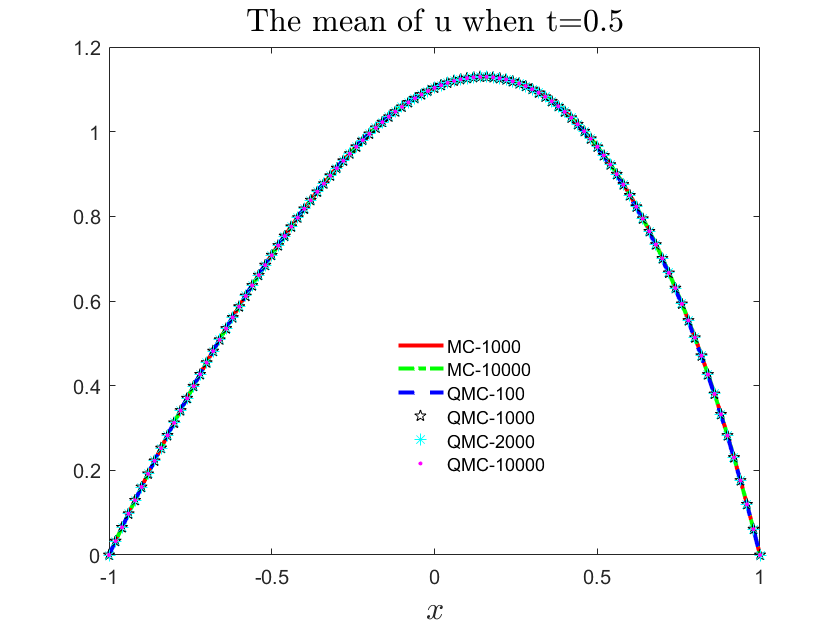}}
\end{minipage}
\hfill
\begin{minipage}[]{0.5 \textwidth}
\centerline{\includegraphics[width=6cm,height=4cm]{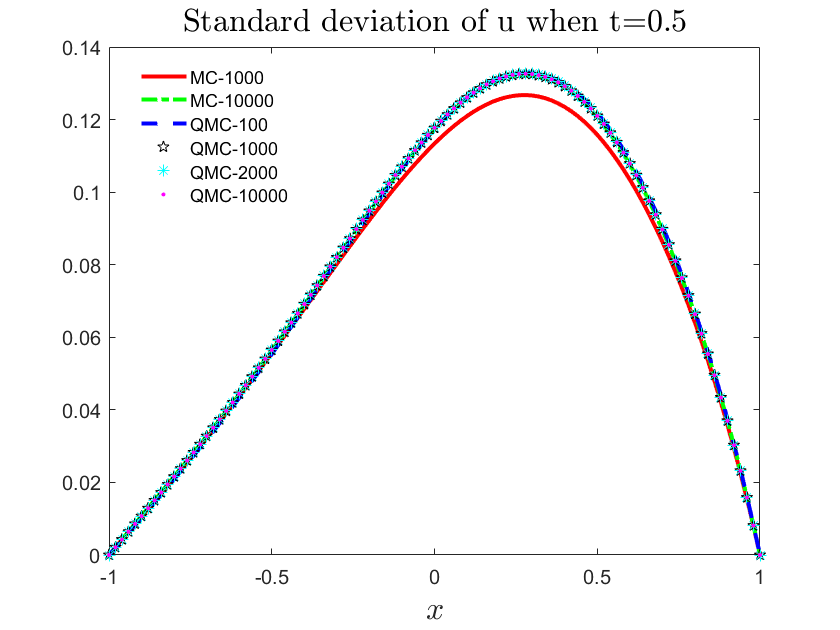}}
\end{minipage}
\begin{minipage}[]{0.5 \textwidth}
\centerline{\includegraphics[width=6cm,height=4cm]{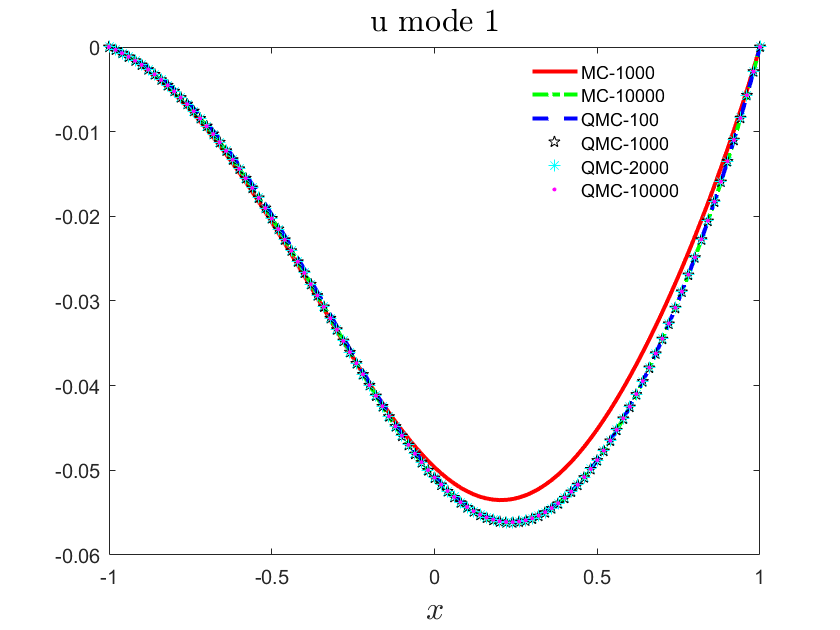}}
\end{minipage}
\hfill
\begin{minipage}[]{0.5 \textwidth}
\centerline{\includegraphics[width=6cm,height=4cm]{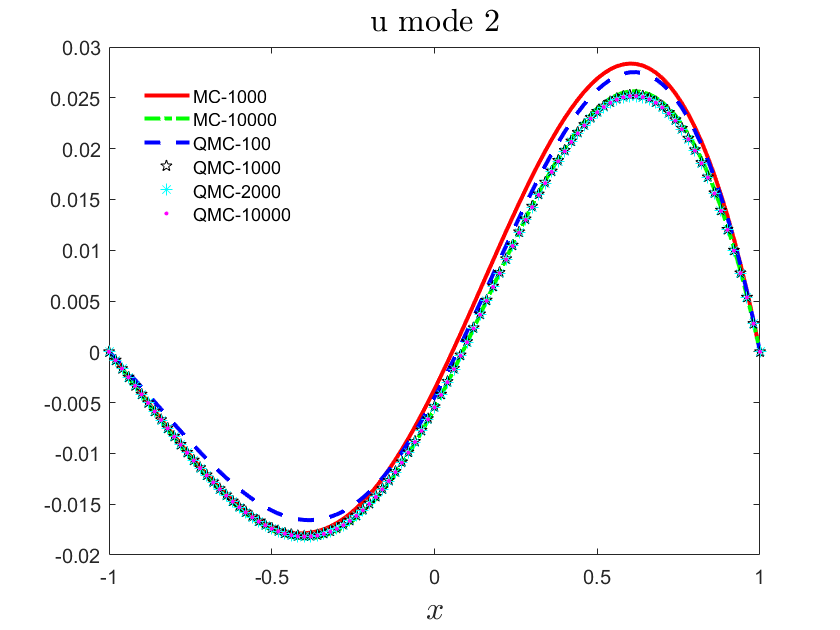}}
\end{minipage}
\hfill
\begin{minipage}[]{0.5 \textwidth}
\centerline{\includegraphics[width=6cm,height=4cm]{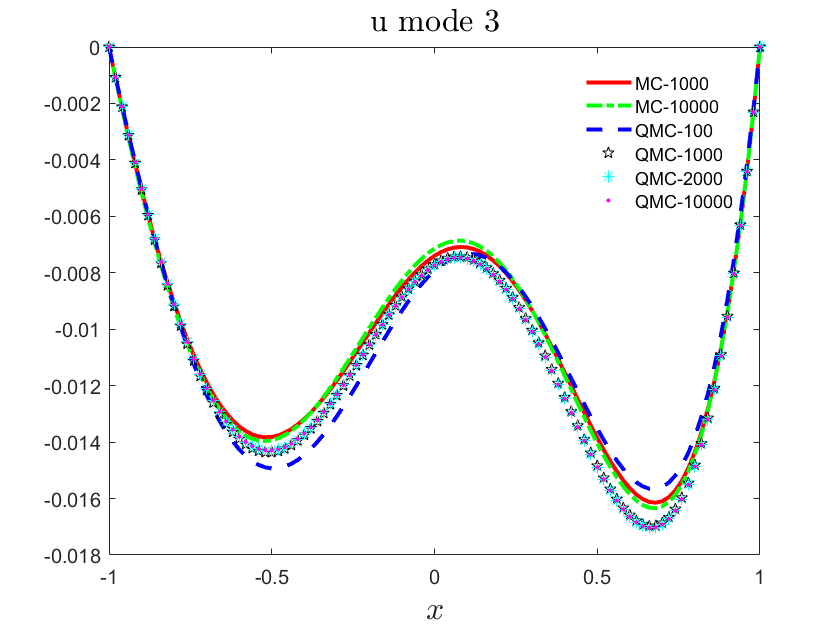}}
\end{minipage}
\begin{minipage}[]{0.5 \textwidth}
\centerline{\includegraphics[width=6cm,height=4cm]{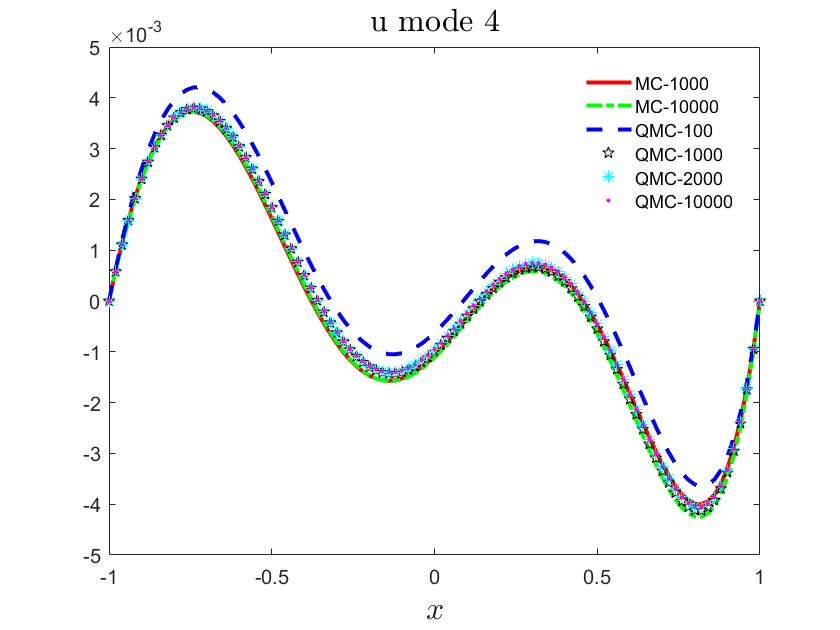}}
\end{minipage}
\caption{\textbf{} The mean, standard deviation and mode functions of u: MC vs QMC.}
\label{fQMC-U1}
\end{figure}

In section \ref{Stoc}, we use the difference method with the Qusi-Monte Carlo method to obtain the reference modes of $k$ and $u$.
In order to estimate how many samples we need for a converged solution, we compare the results with different samples using the Monte Carlo (MC) and the Quasi-Monte Carlo (QMC) methods.
In Figs. \ref{fQMC-K1} and \ref{fQMC-U1}, we present the corresponding results using MC method and QMC method. We can see that the QMC method converges much faster than the MC method. For our examples, we use 2,000 QMC samples for training data, and to obtain the reference solutions we use 10,000 samples.

\bibliographystyle{model1-num-names}
\bibliography{stochastic}







\end{document}